\documentclass[aps,prd,twocolumn,superscriptaddress,floatfix,nofootinbib]{revtex4-2}

\usepackage{amsmath,amssymb,amsfonts}
\usepackage{bm}
\usepackage{graphicx}
\graphicspath{{figs/}{./}}   % figures live in figs/, or flat beside the .tex
\usepackage{hyperref}
\hypersetup{colorlinks=true,linkcolor=blue,citecolor=blue,urlcolor=blue}

\newcommand{\GB}{\mathcal{G}}
\newcommand{\dd}{\mathrm{d}}

\newcommand{\Mpl}{M_{\rm Pl}}
\newcommand{\Mp}{M_{p}}
\newcommand{\lcg}{\ell_{\rm cg}}
\newcommand{\ldef}{\ell_{\rm def}}
\newcommand{\dchidV}{\frac{\partial\chi}{\partial V}}
\newcommand{\Leff}{\Lambda_{\rm eff}}
\newcommand{\CST}{{\rm CST}}
\newcommand{\ab}{\bar\alpha}
\usepackage{orcidlink}

\begin{document}

\title{Topology change and the Gauss--Bonnet term in causal set theory:\\
the sign of the cosmological term and the axion decay constant}

\author{Stylianos A.\ Tsilioukas\,\orcidlink{0009-0003-3051-3405}}
\email{tsilioukas@sch.gr}
\affiliation{\mbox{National Observatory of Athens, Lofos Nymfon, 11852 Athens, 
Greece}}

\date{\today}

\begin{abstract}
Causal set theory permits changes of spatial topology without the metric
degeneracy that forbids them in continuum Lorentzian gravity. In four dimensions
such changes enter the gravitational action only through the Gauss--Bonnet term,
which is the Euler density. We show that the Euler density of a discrete space is
exact, local and combinatorial: it sums to the Euler characteristic at any
resolution and is supported on topological defects. The Gauss--Bonnet term of a
causal set is therefore fixed rather than postulated. Because it is linear in the
number of elements, which is the four-volume, it acts as a cosmological constant.
Its coefficient follows without forming any functional derivative, so the known
obstruction in the continuum does not arise. The everpresent cosmological
constant of causal set theory assumes that the mean topological charge vanishes.
We show that it does not. The species of opposite sign solves the field
equations only if a positive cosmological constant is already present, and,
being simply connected, it cannot carry the axion charge that supports the other
species. Only one sign survives, and the sign of the cosmological term is
derived rather than assumed. We recover the surviving transition from the causal
order alone in every simulated sprinkling. The action of the corresponding axion
wormhole then fixes the magnitude. The observed value requires an axion decay
constant near the grand unification scale, and this result is insensitive to the
one-loop prefactor and to the Gauss--Bonnet coupling.
\end{abstract}

\maketitle

\section{Introduction}

General relativity remains the most successful description of the
gravitational interaction, yet it is incomplete in two respects that point in
opposite directions. Treated as a quantum field theory, it is perturbatively
nonrenormalisable~\cite{tHooft:1974toh, Goroff:1985th}. One alternative route to a consistent
quantum theory of gravity is to abandon the continuum description at short
distances. How that departure is made is the subject of this work.

At the opposite extreme of scale, the concordance cosmological model built on
general relativity explains the cosmic microwave background, primordial
nucleosynthesis and the growth of structure with a small number of
parameters~\cite{Planck:2018vyg}. It does so at the cost of two components whose
microphysical origin is unknown. Dark matter is required by galactic dynamics
and by the acoustic peaks. Dark energy, represented by a cosmological constant,
is required by the late-time acceleration. Together they make up the
overwhelming majority of the present energy budget. The cosmological constant is
also the single worst quantitative discrepancy in physics: its observed value
lies some hundred and twenty orders of magnitude below the natural scale of
vacuum fluctuations~\cite{Weinberg:1988cp}.

These structural problems are now joined by statistical ones. The Hubble
constant inferred from the microwave background within the concordance model
disagrees with the local distance-ladder value at more than five standard
deviations~\cite{Riess:2021jrx,DiValentino:2020zio}. This discrepancy has
resisted explanation in terms of local systematics. A milder but persistent
tension affects the amplitude of matter clustering: weak lensing and large-scale
structure prefer a lower value than the microwave
background~\cite{DiValentino:2020vvd}. Whether these tensions signal new physics
or unmodelled systematics is still open, and the question has been reviewed at
length~\cite{Abdalla:2022yfr,CosmoVerseNetwork:2025alb}.

Proposed responses fall into two broad classes. The first modifies the
gravitational sector. It replaces the Einstein--Hilbert action with a more
general function of the curvature, the torsion or the nonmetricity, and so
obtains an effective dark energy of geometric
origin~\cite{CANTATA:2021asi,Cai:2015emx,Anagnostopoulos:2021ydo}. The second
keeps general relativity. It either introduces a dynamical dark-energy component
or modifies the thermodynamic relation from which the Friedmann equations can be
derived.

Of particular relevance here are scenarios that trace the dark-energy sector to
a quantum-gravitational property of spacetime rather than to an additional
field. Topological dark energy is one of these. In it, the Euler characteristic
of spacetime is allowed to fluctuate. The Gauss--Bonnet term, which is the part
of the gravitational action sensitive to such a change, then generates an
effective cosmological term~\cite{Tsilioukas:2023tdw,Anagnostopoulos:2025tax}.
Related constructions obtain the dark-energy sector from horizon thermodynamics
with a modified entropy, driven by the formation and merger history of black
holes~\cite{Tsilioukas:2024seh,Petronikolou:2025mlm}, or from a deformed
entropy-area law~\cite{Basilakos:2026atu}. Such models have been tested against
early- and late-time data and examined in relation to the tensions
above~\cite{Tsilioukas:2026gvy}. In all of these scenarios the dark-energy
sector is not postulated but inherited from the short-distance structure of
spacetime. The nature of that structure therefore becomes a phenomenological
question, not merely a formal one~\cite{Addazi:2021xuf}.

The weakness of the topological scenarios developed so far is that they are
formulated in the continuum. There the Euler characteristic is an integer
attached to a manifold, and its variation is not a well-defined
operation~\cite{Paschalis:2026top}. This calls for a framework in which
spacetime is discrete from the outset. In such a framework a change of topology
is an ordinary configuration rather than a singular limit, and the Euler
characteristic is a finite combinatorial quantity. Causal set theory provides
both. The rest of this introduction describes what it offers.

Causal set theory~\cite{Bombelli:1987aa,Surya:2019ndm} holds that spacetime is
fundamentally a locally finite partially ordered set. The order relation
represents causal precedence, and counting elements gives four-volume. The
motivation is that the causal structure of a Lorentzian manifold fixes its
metric up to a local conformal factor. Adding a volume measure to the order is
therefore enough to recover the geometry. A manifold approximates a causal set
when the causal set arises from a Poisson sprinkling of fixed density. This
prescription makes the discreteness statistically Lorentz
invariant~\cite{Bombelli:2006nm}. We use the identification of element number
with four-volume throughout. Because the two are the same quantity, any
contribution to the causal-set action proportional to the cardinality is a
cosmological constant term.

Causal sets are a natural setting for topology change. In continuum Lorentzian
gravity, Geroch's theorem~\cite{Geroch:1967fs} forbids a change of spatial
topology within a compact cobordism, unless causality fails or the metric
degenerates. A causal set has no metric that could degenerate: a
topology-changing history is no more than a particular partial order. Sorkin has
argued that topology change is required for a consistent quantum theory of
gravity~\cite{Sorkin:1997gi,Sorkin:1986cs}, with the allowed transitions
constrained by causal continuity~\cite{Dowker:1997hj}. Recovering topological
information from a causal set is nevertheless delicate. The spatial homology of
an inextendible antichain can be extracted from the nerve of a thickening of
that antichain~\cite{Major:2006hv,Major:2009cw}, but only within a stability
window. Topological quantities in a causal set therefore carry an explicit
resolution scale.

If the topology fluctuates, the fluctuation enters the gravitational action
through the Gauss--Bonnet invariant, which in four dimensions is the Euler
density~\cite{Chern:1945aa,Eguchi:1980jx}. Classically this density is a total
derivative and carries no local dynamics. Its entire content is topological. It
is therefore the unique curvature term that responds to a change of topology and
to nothing else. Any effect of spacetime
foam~\cite{Wheeler:1957mu,Hawking:1977jb} on the gravitational action must pass
through it.

Causal set theory has its own prediction for the cosmological constant, and it
is the best-known phenomenological result of the programme. Because $\Lambda$
and the four-volume are conjugate variables, discreteness makes the four-volume
fluctuate by an amount of order its square root. The magnitude of $\Lambda$ is
then comparable to the critical density at every epoch. This is the everpresent
$\Lambda$ of Sorkin~\cite{Sorkin:1990bh,Ahmed:2002mj}, which anticipated the
observed order of magnitude. The cosmological viability of the resulting
stochastic models has been examined in detail~\cite{Zwane:2017xbg}.

The mechanism we study is the first moment of the same underlying process. A
foam of topology-changing events adds to the Euler characteristic a sum of
integer increments. The variance of that sum reproduces the everpresent scaling,
while the mean contributes a constant. The standard treatment takes the mean to
vanish, as it must if species of both signs nucleate at comparable rates. Our
central dynamical result is that they do not, because only one species admits
matter support. The mean therefore survives, is constant, and can be calculated.
The two channels are the first two moments of a single Poisson process and
coexist. Nothing established below bears on the viability of the fluctuation
channel, which we do not attempt to settle.

We now return to the topological dark energy scenario introduced above. Its
continuum formulation~\cite{Tsilioukas:2023tdw,Anagnostopoulos:2025tax} obtains
the effective cosmological term by semiclassical variation of the
Einstein--Gauss--Bonnet action on a topologically fluctuating background. The
resulting term is proportional to the derivative of the Euler characteristic
with respect to the four-volume. In that work the density of topology-change
events was estimated from Euclidean instantons, which left the magnitude
undetermined. We recover the same expression within causal set theory. Here it
follows without forming any functional derivative, and its magnitude can be
calculated. Throughout, we keep the Gauss--Bonnet coupling as a free parameter
and take it to be positive. Together with the results of Sec.~\ref{sec:flow},
this fixes the sign of the effective term.

The paper is organised as follows. In Sec.~\ref{sec:kappa} we show that the
Euler density of a discrete space is an exact, local, combinatorial quantity,
and we identify what a bare partial order does and does not provide. In
Sec.~\ref{sec:topo} we treat topology change and report numerical
demonstrations of the discriminating observable in $2+1$ and $3+1$ dimensions;
the $3+1$-dimensional demonstration recovers the wormhole-sector transition from
the causal order alone. In Sec.~\ref{sec:derive} we construct the causal-set
Gauss--Bonnet term and derive the effective cosmological term from its linearity
in the element count. In Sec.~\ref{sec:flow} we determine which
topology-changing species are available, compute the handle action, and derive
the resulting prediction for the axion decay constant. Section~\ref{sec:renorm}
draws the consequences for renormalisation and cosmology, and
Sec.~\ref{sec:disc} summarises what has and has not been established.

\section{The Euler density of a discrete space}
\label{sec:kappa}

In four dimensions the Gauss--Bonnet invariant
$\GB=R^{2}-4R_{\mu\nu}R^{\mu\nu}+R_{\mu\nu\rho\sigma}R^{\mu\nu\rho\sigma}$ is the
Euler density. For a closed manifold $M$,
\begin{equation}
\frac{1}{32\pi^{2}}\int_M\!\dd^4x\sqrt{g}\,\GB=\chi(M).
\label{eq:cgb}
\end{equation}
Since $\chi$ takes integer values, the derivative $\partial\chi/\partial V$ in
the effective term of Refs.~\cite{Tsilioukas:2023tdw,%
Anagnostopoulos:2025tax} cannot be a derivative in the classical sense. It is
locally constant, with jumps. In this section we identify the discrete object of
which \eqref{eq:cgb} is the coarse-grained statement, and we establish which of
its properties survive in a bare partial order.

\subsection{The combinatorial density and its properties}
\label{sec:regge}

For any finite simplicial complex we define
\begin{equation}
\kappa(v)\;=\;\sum_{k\ge0}\frac{(-1)^{k}\,n_k(v)}{k+1},
\label{eq:kappa}
\end{equation}
where $n_k(v)$ is the number of $k$-simplices containing the vertex $v$. A
$k$-simplex has $k+1$ vertices, so distributing the alternating face count over
the vertices gives
\begin{equation}
\sum_{v}\kappa(v)=\sum_k(-1)^{k}N_k=\chi
\label{eq:localGB}
\end{equation}
identically. Equation~\eqref{eq:localGB} is a combinatorial Gauss--Bonnet
theorem, and $\kappa$ is the unique locally determined numerical homotopy
invariant~\cite{Levitt:1992aa}. The relation \eqref{eq:localGB} is an identity, not a limit. It
holds for every finite complex and at every mesh, with no manifold and no
refinement required. Throughout this work $\kappa$ denotes the density
\eqref{eq:kappa}. It never denotes the gravitational coupling, for which we
write $8\pi G$ explicitly.

On a triangulated surface, a vertex of degree $d$ lies in $d$ edges and $d$
triangles, which gives $\kappa(v)=1-d/6$. On an equilateral triangulation every
angle equals $\pi/3$. There $\kappa$ coincides with the Regge angle
deficit~\cite{Regge:1961px} divided by $2\pi$,
\begin{equation}
\kappa(v)=\frac{1}{2\pi}\Big(2\pi-\tfrac{\pi}{3}d\Big),
\label{eq:regge}
\end{equation}
where the normalisation makes $\sum_v\kappa=\chi$ rather than $2\pi\chi$.

The restriction to equilateral triangulations in \eqref{eq:regge} matters,
because away from that case the two quantities are different objects. The Regge
deficit is metric: it depends on the angles, and therefore on the embedding. By
contrast, \eqref{eq:kappa} is purely combinatorial and depends only on the
degree. Both sum to the Euler characteristic on any triangulation, since
$\sum_v\delta(v)=2\pi\chi$ and $\sum_v\kappa(v)=\chi$, but they need not agree
pointwise. On a Delaunay triangulation of randomly sampled points on $S^{2}$ we
find them anticorrelated, with $\max_v|\kappa-\delta/2\pi|$ of order unity. The
difference is sharpest under refinement. On a repeatedly subdivided icosahedron,
$\sum_v\kappa=2$ exactly at every level, while the density stays on the twelve
original vertices. The deficit $\delta$, in contrast, spreads over all vertices
and approaches the smooth Gaussian curvature density, as the convergence theorem
for piecewise flat spaces requires~\cite{Cheeger:1984zz}.

The metric independence of $\kappa$ is not a deficiency. It is the reason
$\kappa$ is the appropriate object here. The Gauss--Bonnet term is topological,
and a density for it must not depend on the metric. In every case the mechanism
developed below needs only the total, which \eqref{eq:localGB} supplies exactly
and at every resolution. We have verified three further properties numerically.

First, exactness. On the octahedron, on the $3\times3$ flat torus and on the
filled triangle, Eq.~\eqref{eq:localGB} returns $\chi=+2$, $0$ and $+1$ exactly.
Within the stable window of an alpha-complex filtration on Poisson samples of
$S^{2}$ with $n=400$ and of $T^{2}$ with $n=700$, the sum $\sum_v\kappa$ returns
$2.000$ and $0.000$, with vanishing scatter over six samples.

Second, flat regions carry no density. On a $6\times6$ triangulated flat torus
we find $\max_v|\kappa(v)|=4.4\times10^{-16}$. The Euler density therefore
vanishes pointwise, not merely in the aggregate.

Third, defects carry the density atomically. A single edge flip produces two
vertices of degree five and two of degree seven, with $\kappa=\pm1/6$, while
$|\kappa|<10^{-15}$ elsewhere. The sum remains $\chi=0$: the flip is a curvature
dipole, not a change of topology.

These properties carry over unchanged to four dimensions. For
$S^{4}=\partial\Delta^{5}$ we obtain $\sum_v\kappa=+2.000000$, with $\kappa=1/3$
at each of the six vertices. For a Kuhn-triangulated flat four-torus with
$3^{4}=81$ vertices and $1944$ four-simplices, $\sum_v\kappa=1.5\times10^{-13}$
and $\max_v|\kappa(v)|=1.9\times10^{-14}$.

\subsection{The atomic measure, and what a causal set provides}
\label{sec:complex}

The results above identify the discrete counterpart of the Euler density as an
atomic measure,
\begin{equation}
\frac{1}{32\pi^{2}}\,\GB(x)\ \longleftrightarrow\ \mu_\chi
=\sum_i\delta\chi_i\,\delta^{(4)}(x-x_i),
\label{eq:atomic}
\end{equation}
which is a property of discreteness rather than an assumption about foam. The
correspondence \eqref{eq:atomic} should be read as an equality of integrals:
integrating it returns \eqref{eq:cgb}, which at the discrete level is an
identity rather than a theorem to be recovered. We do not claim that $\kappa$
converges pointwise to the smooth Euler density, and the icosahedron of
Sec.~\ref{sec:regge} shows that it does not. Continuum curvature can emerge from
a distribution of disclinations, and a smooth curvature might emerge from a
cloud of defects in the same way. That, however, is a statement about the metric
curvatures~\cite{Cheeger:1984zz}, and it requires a nondegeneracy condition on
the simplices. The complexes constructed in
Secs.~\ref{sec:topo}--\ref{sec:derive} do not manifestly satisfy that condition.
Nothing in what follows depends on it.

We split the Euler characteristic of a region by resolution,
\begin{equation}
\chi=\underbrace{\frac{1}{32\pi^{2}}\!\int\!\sqrt{g}\,\GB_{\rm smooth}}_{\text{background}}
\;+\;\underbrace{\sum_i\delta\chi_i}_{\text{foam defects}} ,
\label{eq:split}
\end{equation}
where, for a Friedmann--Robertson--Walker background,
$\GB_{\rm smooth}=24H^{2}(\dot H+H^{2})$ is a total derivative that carries no
dynamics. The defect sum in \eqref{eq:split} is a count of point events. It is
therefore exactly additive over any decomposition of the region, with no
boundary or corner terms. The derivation of Sec.~\ref{sec:Nlinear} rests on this
property.

Two features of \eqref{eq:split} need comment. First, the increments
$\delta\chi_i$ are increments relative to the background, not a decomposition of
it. Attaching a handle $X$ by connected sum changes the Euler characteristic by
$\delta\chi=\chi(X)-2$. This gives $-2$ for $S^{1}\!\times\!S^{3}$ and $+2$ for
$S^{2}\!\times\!S^{2}$, as in \eqref{eq:dchi} below. The sum $\sum_v\kappa(v)$
over all vertices returns the total, not the background value. The asymmetry
between the two terms of \eqref{eq:split} is the entire mechanism: the first
does not grow with the four-volume, while the second does. For a region rather
than a closed manifold, the first term is not an integer on its own. The
Chern--Gauss--Bonnet theorem then carries a boundary contribution, which must be
added to recover $\chi(\Omega)$. Physically, what matters is only that this term
as a whole is subextensive: it scales as the boundary three-volume, not as the
four-volume.

Second, the split depends on resolution by construction. What counts as a defect
rather than as background curvature depends on the scale at which the causal set
is coarse grained. Section~\ref{sec:threshold} studies that dependence, and
Sec.~\ref{sec:species} shows that it drops out of the final prediction.

Equation~\eqref{eq:kappa} presupposes a simplicial complex, whereas a causal set
is a bare partial order. We tested three purely order-theoretic candidates, and
all three fail to reproduce the topology of the approximated manifold. The order
complex returns $\chi=+1$ for any poset with a maximum, since its order complex
is then a contractible cone. Otherwise it diverges. For sprinklings of a
$1{+}1$-dimensional causal diamond it grows from $\{+3,+1,+2,+1\}$ at $N=20$ to
$\{-154,+38,+33,+43\}$ at $N=160$. For a cylinder slab of vanishing Euler
characteristic it scatters between $-28$ and $+174$. The nerve of inclusive
future cones is likewise trivial for a diamond and fails to converge for the
cylinder. A cover by bounded Alexandrov intervals scatters between $-7$ and $+5$
for the diamond, and between $-10$ and $-4$ for the cylinder. These failures
have a structural cause. The nerve theorem requires a good cover, but the natural
sets in Lorentzian signature are causal cones. These are unbounded and overlap
globally, and their intersections are generically disconnected precisely when
the spatial topology is nontrivial. These tests are limited to $N\lesssim20$ by
the exhaustive subset enumeration they require, so finite-size effects cannot be
excluded.

We therefore adopt the spatial route of Refs.~\cite{Major:2006hv,Major:2009cw},
together with the Morse construction of Sec.~\ref{sec:bridge}. We regard the
construction of an order-theoretic complex that reproduces four-dimensional
spacetime topology as an open problem of causal set theory. The causal-set Euler
characteristic is therefore defined but not directly computable, a limitation to
which we return in Sec.~\ref{sec:disc}. An independent route through
higher-curvature operators is at present closed.
Reference~\cite{Pfeiffer:2023hcs} constructs discrete operators that converge to
$(\Box-R/2)^{n}$, and hence to $R^{2}-2\Box R$, but only within the Ricci-scalar
sector. Operators converging to $R_{\mu\nu}R^{\mu\nu}$ and to
$R_{\mu\nu\alpha\beta}R^{\mu\nu\alpha\beta}$ are listed there as open, because
objects with open indices require a notion of tangent space. The Gauss--Bonnet
combination needs exactly those two invariants. It therefore cannot at present
be assembled from such operators, and $\kappa$ remains the only available route
to the Euler term.

\section{Topology change and its detection from the causal order}
\label{sec:topo}

\subsection{From spatial surgery to spacetime handles}
\label{sec:bridge}
\label{sec:betti}

A moment of time in a causal set is an inextendible antichain. We thicken such
an antichain and form the nerve of the resulting cover. This yields spatial Betti
numbers, which are meaningful within a stability window in the thickening
parameter. On Poisson samples of two-dimensional surfaces the procedure is
reliable. Alpha-complex persistent homology returns $(1,0,1)$ with $\chi=+2$ for
$S^{2}$, and $(1,2,1)$ with $\chi=0$ for $T^{2}$, in $12/12$ samples.

In $3+1$ dimensions this route is closed at the outset, because every closed
three-manifold has $\chi(\Sigma^{3})\equiv0$. The spatial Euler characteristic
therefore carries no information whatever. The four-dimensional $\delta\chi$
must instead be treated as a bulk quantity. The spatial transition of interest
is the index-one handle attachment and its dual,
\begin{equation}
S^{3}\ \xrightarrow{\ \lambda=1\ }\ S^{1}\!\times\!S^{2}
\ \xrightarrow{\ \lambda=3\ }\ S^{3},
\label{eq:spatialseq}
\end{equation}
whose associated four-dimensional topology is $S^{1}\!\times\!S^{3}$, the closed
manifold obtained by capping the cobordism, and not the topology of any single
slice. The bridge from the sequence of slices to the bulk invariant is
Morse-theoretic~\cite{Sorkin:1986cs,Dowker:1997hj}:
\begin{equation}
\chi(W)=\chi(\Sigma_0)+\sum_i(-1)^{\lambda_i}
\xrightarrow[\chi(\Sigma_0)=0]{}
\chi(W)=\sum_i(-1)^{\lambda_i},
\label{eq:morse}
\end{equation}
and the Euler characteristic of a four-volume is therefore the signed count of
the topology-change events it contains. For the cobordism carrying the handle
pair, which is $S^{1}\!\times\!S^{3}$ with two four-balls removed, this gives
$\chi(W)=-2$. Capping the two boundaries restores $\chi(S^{1}\!\times\!S^{3})=0$.
We have verified the parity of \eqref{eq:morse} on the $1+1$-dimensional pair of
pants, whose spatial slices are circles with $\chi=0$. An exact triangulation
returns $(1,2,0)$ and $\chi(W)=-1$, matching the single index-one saddle. The two
species relevant below are accordingly
\begin{align}
\text{wormhole sector}:&\quad \delta\chi=-2,\nonumber\\
\text{Nariai sector }(S^{2}\!\times\!S^{2}):&\quad \delta\chi=+2,
\label{eq:dchi}
\end{align}
with topological density
\begin{equation}
\dchidV=\sum_i \delta\chi_i\,n_i .
\label{eq:density}
\end{equation}
Which species is realised is a dynamical question, and the sign of $\Leff$
depends on the answer. Section~\ref{sec:nariai} settles it.

% ===================================================================
% FIG 1 -- topology change in a sprinkled causal set (schematic)
% ===================================================================
\begin{figure}[t]
\centering
\includegraphics{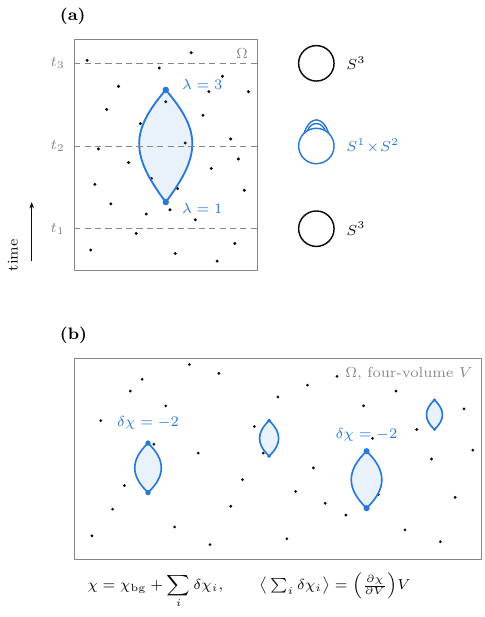}
\caption{Topology change in a causal set. \emph{(a)} A single wormhole-sector
event inside a region $\Omega$. The shaded lens is the worldvolume of the handle.
It is bounded below and above by the two Morse critical points of index
$\lambda=1$ and $\lambda=3$ identified in Eq.~\eqref{eq:spatialseq}. The spatial
slice at the three indicated moments of time, each an inextendible antichain,
passes from $S^{3}$ to $S^{1}\!\times\!S^{2}$ and back. By \eqref{eq:morse} the
event therefore carries $\delta\chi=(-1)^{1}+(-1)^{3}=-2$. Dots are elements of
the causal set. The sprinkling, rather than a lattice, is what makes the
discreteness Lorentz invariant. \emph{(b)} A foam of such events, of varying
size, occurring at a fixed rate per unit four-volume. The Euler characteristic of
the region is the background value plus an exactly additive integer count. The
extensivity of that count, $\langle N_i\rangle=n_iV$, makes $I_{GB}$ linear in
the cardinality and hence a cosmological constant term, as shown in
Sec.~\ref{sec:Nlinear}.}
\label{fig:topochange}
\end{figure}

The Betti vectors of the two slice topologies,
$\beta(S^{3})=(1,0,0,1)$ and $\beta(S^{1}\!\times\!S^{2})=(1,1,1,1)$, give
\begin{equation}
\Delta\bm\beta_{\rm create}=(0,+1,+1,0),\qquad
\Delta\bm\beta_{\rm destroy}=(0,-1,-1,0),
\label{eq:dbeta}
\end{equation}
which identify the Morse indices $\lambda=1,3$. Hence they yield
$\delta\chi=-2$ from data that are, in principle, extractable from the order
alone. The identification is necessary but not sufficient. The wedge
$S^{1}\vee S^{2}\vee S^{3}$ carries the same Betti vector, and the two spaces are
distinguished only by the cup product. We have constructed an explicit
seven-vertex complex with $\beta=(1,1,1,1)$, namely $\partial\Delta^{4}$ with one
vertex generating a $1$-cycle and one a $2$-cycle. It is a wedge and not a
three-manifold. The manifold condition must therefore be imposed separately, and
to that extent the extraction of $\delta\chi$ from order data is not rigorous. In
$3+1$ dimensions, where $\chi(\Sigma^{3})\equiv0$ removes the spatial route
entirely, \eqref{eq:morse} is the only bridge available. The Betti vector itself
is nevertheless a measurable quantity, and we measure it in the remainder of this
section.

\subsection{Detection in $2+1$ dimensions}
\label{sec:numerics}

The construction above is a prescription for extracting topology change from the
order alone. As such, it can be tested. We first report an end-to-end
demonstration in $2+1$ dimensions. There the spatial slice is a surface and $b_1$
is directly meaningful. The parity obstruction $\chi(\Sigma^{3})\equiv0$ of the
physical case is absent, and the observable can therefore be checked against a
known answer.

\emph{Geometry, and what is specified rather than derived.} By Geroch's theorem,
no smooth Lorentzian trousers metric exists. There is therefore nothing to
sprinkle in the usual sense, and we specify the causal order directly. This is
not a shortcut. It is precisely the sense in which a causal set admits topology
change, and the object of the exercise is to make it explicit. Away from the
transition the order is that of an ultrastatic cobordism
$\dd s^{2}=-\dd t^{2}+\dd l^{2}$. Its spatial slice is one circle for
$t<t_{\rm split}$ and two thereafter. We set $i\prec j$ if and only if
$t_j-t_i>d(i,j)$, where $d$ is arclength along the slice, and we declare elements
on causally disconnected legs unrelated. The embedding is then discarded, and
every subsequent step uses only the order. A cylinder, whose slice is always a
single circle, serves as a control. We do not claim that the resulting poset is
manifoldlike in a neighbourhood of the transition, since no such statement is
available. What is tested is therefore the observable, not the existence of the
history.

\emph{Pipeline.} We grow an inextendible antichain $A$ greedily from a thin time
band and compute the causal layers above it. We then form the nerve from the
past-shadows $\sigma(y)=\{a\in A:a\prec y\}$ of the elements above $A$, and we
obtain the Betti numbers over $\mathbb{Z}_2$ by rank reduction. We validated the
implementation against the point, interval, circle, disk, octahedral $S^{2}$,
$3\times3$ torus, and disjoint-circle complexes.

\emph{The resolution parameter.} The choice of resolution is what makes the
construction work. Thickening by layer count fails. An element sufficiently far
to the future of $A$ has essentially all of $A$ in its past. Its shadow then
spans the whole antichain, the nerve degenerates to a single large simplex, and
all topology is lost. The correct discrete analogue of a small ball is a shadow
of small cardinality. We therefore retain only witnesses with $|\sigma(y)|\le c$
and scan the resolution $c$. This is the stability window described above, in
order-theoretic form.

\emph{Results.} For $N=8000$ elements the Betti vector $(b_0,b_1)$ as a function
of $c$ is
\begin{center}
\begin{tabular}{lcccccc}
\hline\hline
$c$ & 4 & 6 & 8 & 10 & 12 & 14\\
\hline
cylinder & $(26,0)$ & $(3,0)$ & $(1,1)$ & $(1,1)$ & $(1,1)$ & n/a\\
waist & $(6,0)$ & $(1,1)$ & $(1,1)$ & $(1,1)$ & $(1,1)$ & n/a\\
legs & $(30,0)$ & $(2,1)$ & $(2,2)$ & $(2,2)$ & $(2,2)$ & n/a\\
\hline\hline
\end{tabular}
\end{center}
where n/a denotes a nerve too large to reduce. Small $c$ under-resolves and
fragments the slice into spurious components. Large $c$ over-connects. Between
the two lies a plateau, $c\simeq8$ to $12$, on which the correct answer is
returned: one circle at the waist and two at the legs. The transition
\begin{equation}
(b_0,b_1):\quad (1,1)\ \longrightarrow\ (2,2)
\label{eq:2p1transition}
\end{equation}
is thus recovered from the causal order alone. Over eight independent
sprinklings at $N=8000$ and $c=10$, the detection rate is $8/8$.

\emph{Cost.} With a bit-packed causal matrix and a light-cone time window,
building the order and extracting the topology take
\begin{center}
\begin{tabular}{lcccc}
\hline\hline
$N$ & $3000$ & $6000$ & $12000$ & $24000$\\
$|A|$ & $84$ & $146$ & $196$ & $275$\\
build & $0.1\,$s & $0.3\,$s & $1.0\,$s & $4.2\,$s\\
topology & $0.0\,$s & $0.1\,$s & $0.2\,$s & $0.5\,$s\\
\hline\hline
\end{tabular}
\end{center}
on a single core. The build scales as $N^{2}$ and the topology as $|A|$.

These results establish three points. Spatial topology change is detectable from
the order alone by the construction used here. A stability plateau exists. The
resolution must be fixed by shadow cardinality. The physical case is $3+1$
dimensional, and we turn to it now.

\subsection{Detection in $3+1$ dimensions}
\label{sec:numerics4}

The wormhole sector of \eqref{eq:spatialseq} is an index-one surgery
$S^{3}\to S^{1}\!\times\!S^{2}$. The same pipeline applies to it once two
corrections are made. Transferring the $2+1$-dimensional recipe unchanged does
not work, and the reason is instructive rather than accidental.

\emph{The resolution is fixed by the dimension of the slice.} A Dowker complex
built on a point sample of a $d$-manifold reproduces the topology only when the
cover sets are balls of radius $r$ equal to a small multiple of the sample
spacing $s$. The number of sampled points in such a ball is
\begin{equation}
m=\omega_d\,(r/s)^{d},\qquad \omega_2=\pi,\quad\omega_3=\tfrac43\pi,
\label{eq:capdim}
\end{equation}
with $\omega_d$ the volume of the unit $d$-ball. The cover cardinality that
resolves a slice is therefore set by the dimension of the slice, not by the
sprinkling density. For $r/s\simeq1.5$, Eq.~\eqref{eq:capdim} gives $m\simeq7$
in two dimensions and $m\simeq14$ in three. This agrees with the plateaux we
observe: $c\simeq8$ to $12$ for the surface slices of Sec.~\ref{sec:numerics},
and $c\gtrsim10$ with $c=14$ clean for the three-dimensional slices below. Run at
$c\lesssim8$, a three-dimensional slice is under-resolved. It returns $b_0$
between $2$ and $10^{2}$ with $b_1$ of order $10^{2}$, a spurious plateau of
discreteness artefacts rather than a topology.

This fragmentation is a diagnostic of a failed reconstruction, not a measurement
that propagates anywhere. The slice is known to be connected, so its correct
value is $b_0=1$. A result $b_0\gg1$ means that the reconstruction has broken
the slice into many spurious pieces. This happens when the cover radius falls
below the sample spacing: neighbouring cover sets then fail to overlap, and the
nerve no longer approximates the slice at all. The appropriate response is to
reject the parameter choice, which is what the plateau criterion does. The
quantity carries no uncertainty into the remainder of the manuscript, because no
measured Betti number enters the effective cosmological term.

The chain of inference runs in the opposite direction.
Equation~\eqref{eq:Leffderived} fixes $\Leff$ in terms of $\dchidV$. By
\eqref{eq:density}, the topological density is a sum of integer increments $\delta\chi_i$
weighted by nucleation rates $n_i$. The Morse relation \eqref{eq:morse} supplies
the increments as exact signed counts of critical points, and the instanton
actions of Sec.~\ref{sec:flow} supply the rates. Neither is obtained by measuring
the homology of a sprinkled slice. The numerical work establishes that the
transition postulated by \eqref{eq:spatialseq} is in fact recoverable from a
causal order. It thus shows that the species entering \eqref{eq:density} is
realisable in a discrete setting. It is an existence demonstration, not a
measurement feeding a subsequent calculation. The same remark applies to
$\chi_\CST$ of \eqref{eq:IGB}, which is defined by the combinatorial density
\eqref{eq:kappa} and is not extracted from any filtration. Alpha-complex methods
appear in this work only as independent validation on two-dimensional surfaces,
where the answer is known in closed form. In $3+1$ dimensions the spatial Euler
characteristic vanishes identically and could not serve as an observable in any
case.

\emph{The complex must be built on a sparse landmark subset.}
Equation~\eqref{eq:capdim} fixes the facet size but not the vertex count. The
vertex count governs the linear algebra, through the dimension $n_1-n_0+b_0$ of
the cycle space in which the triangle boundaries must be reduced. We therefore
take a random subset $L\subset A$ of the antichain as the vertex set and retain
all of $A$ as witnesses, with facets $\sigma(y)\cap L$. The construction remains
purely order-theoretic, since a random subset introduces no metric notion. It is
legitimate by Dowker's theorem~\cite{Dowker:1952aa}. The nerve of a cover and the
complex on the covered set generated by the cover sets have isomorphic homology,
so either side of the relation may be used. Below, $|L|=250$ against
$|A|\simeq900$ to $2200$.

\emph{Geometry.} As in Sec.~\ref{sec:numerics}, we specify the order directly.
The pre-surgery slice is a round $S^{3}$. The post-surgery slice opens the two
poles into $2$-spheres of throat radius $\lambda_{\rm th}$, inserts a cylindrical tube of
length $\ell_{\rm neck}$ at each end, and identifies them. This gives
$S^{1}\!\times\!S^{2}$ with a $1$-cycle of length $2\ell_{\rm neck}+\pi$. Static
slabs with slices $S^{3}$ and $S^{1}\!\times\!S^{2}$, both of which admit an
exact closed-form geodesic distance, serve as controls. A warped-coordinate
$S^{3}$ validates the product approximation used for the surgery metric.

\emph{Results.} For $N=2\times10^{4}$, $|L|=250$, and eight independent
sprinklings, the static controls are recovered as
\begin{center}
\begin{tabular}{lcccc}
\hline\hline
 & $c=8$ & $c=10$ & $c=12$ & $c=14$\\
\hline
$S^{3}\to(1,0)$                & $0/8$ & $7/8$ & $7/8$ & $8/8$\\
$S^{1}\!\times\!S^{2}\to(1,1)$ & $5/8$ & $8/8$ & $8/8$ & $8/8$\\
\hline\hline
\end{tabular}
\end{center}
and at $c=14$ the cobordism, with $\lambda_{\rm th}=0.40$ and $\ell_{\rm neck}=1.5$,
returns $(1,0)$ before the transition and $(1,1)$ after it in every sprinkling:
\begin{equation}
\Delta b_1=+1\quad\text{in }8/8,\qquad N=2\times10^{4}.
\label{eq:db1}
\end{equation}
Detection is already $4/4$ at $N=4\times10^{3}$ and does not degrade upward. The
antichain grows as $|A|\propto N^{0.743}$, consistent with the expected
$N^{3/4}$. The cost is $5.1\,$s per sprinkling at $N=2\times10^{4}$ and $63\,$s
at $N=1.6\times10^{5}$ on a single core, in under $0.1\,$GB. The transition is
therefore detectable in $3+1$ dimensions from the causal order alone. This holds
at a cardinality an order of magnitude below the $5\times10^{4}$ associated with
the dominant wormhole.

% ===================================================================
% FIG 2 -- the 3+1D numerics: plateau, discrimination, and the residual
% ===================================================================
\begin{figure*}[t]
\centering
\includegraphics{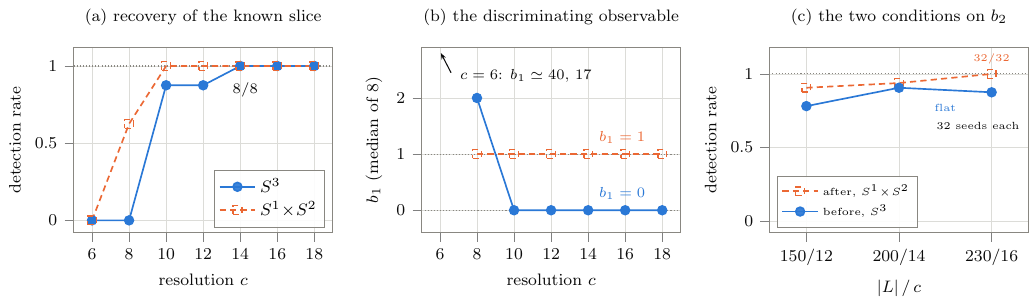}
\caption{Recovering three-dimensional slice topology from the causal order alone,
at $N=2\times10^{4}$ with $|L|=250$ landmarks and eight independent sprinklings.
Marker shape and dash pattern carry identity as well as colour, and the panels
therefore survive greyscale reproduction. \emph{(a)} Fraction of sprinklings
returning the correct $(b_0,b_1)$ against the shadow-cardinality resolution $c$.
Below $c\simeq8$ the three-dimensional slice is under-resolved, as
Eq.~\eqref{eq:capdim} requires. From $c=14$ to $c=18$ both manifolds are
recovered in every sprinkling. \emph{(b)} The observable itself: $S^{3}$ settles
on $b_1=0$ and $S^{1}\!\times\!S^{2}$ on $b_1=1$, and the two are cleanly
separated across the whole plateau. At $c=6$ the medians are $40$ and $17$, off
scale. This is a spurious plateau of discreteness artefacts, of the kind produced
by a resolution transferred unchanged from the two-dimensional case. \emph{(c)}
The two conditions governing $b_2$, which admits no cycle-space shortcut and must
be obtained from the three-simplices. The panel shows detection rates over $32$
sprinklings at $N=2\times10^{5}$ for three settings of the landmark count and the
resolution. We chose the settings by \eqref{eq:twoknobs} to vary the cover radius
and the solidity separately. The cover radius limits the post-transition slice,
since a cover that is too large fills the throat's $2$-cycle. That slice reaches
$32/32$ once both quantities are satisfied. The pre-transition slice is flat
across the three settings. Its failures are spurious voids that do not respond to
$c$. They are instead the signature of a slice witnessed some five times less
densely than its partner at equal $|L|$, the two differing in volume by a factor
of three and a half. The residual is therefore undersampling rather than
under-resolution.}
\label{fig:numerics}
\end{figure*}

\emph{The full Betti vector, and the two conditions it imposes.} Recovering
$b_2$ is harder than recovering $b_1$, because it constrains two independent
quantities rather than one. A witness at height $\eta$ above the antichain has a
shadow containing roughly $\rho_L(4\pi/3)\eta^{3}$ landmarks. The window
admitting shadows of at most $c$ landmarks therefore has
\begin{equation}
\eta=\Big(\frac{3cV_\Sigma}{4\pi|L|}\Big)^{1/3},\qquad
\frac{\eta}{s}=\Big(\frac{3c}{4\pi}\Big)^{1/3},
\label{eq:twoknobs}
\end{equation}
with $V_\Sigma$ the volume of the slice and $s=(V_\Sigma/|L|)^{1/3}$ the
landmark spacing. The ratio $\eta/s$ governs how solid the complex is in
dimension three, and it depends on $c$ alone: the landmark count cancels. The
cover radius $\eta$ itself depends on $c/|L|$. It decides whether a real feature
of a given physical size survives or is filled in. At fixed $c$ the two
requirements pull in opposite directions, which is why no single parameter
suffices.

Three distinct failure modes follow, each attached to one of these quantities,
and we observe all three. First, when $\eta$ exceeds the throat radius, the
genuine $2$-cycle of the post-transition slice is filled. The slice is then
returned as $\bm\beta=(1,1,0)$, a solid torus. This occurs stably rather than
erratically: at fixed $c=12$, $|L|=120$ produces it in half of all sprinklings.
Second, when $\eta/s$ is too small, the complex is ragged and spurious voids
appear. These raise $b_2$ above its true value on either slice. Third, when the
number of witnesses $\sim V_\Sigma\,\eta$ is small, the nerve is assembled from
too few facets. The same spurious voids then appear, for a reason that no
increase in $c$ can remedy. Reducing $\eta$ eliminates the first mode, and
raising $c$ eliminates the second. Neither change addresses the third.

Detection rates at three settings, each over $32$ sprinklings at
$N=2\times10^{5}$, $\lambda_{\rm th}=0.80$ and $\ell_{\rm neck}=1.5$, are
\begin{center}
\begin{tabular}{ccccccc}
\hline\hline
$|L|$ & $c$ & $\eta/s$ & $\eta$ & before & after & both\\
\hline
$150$ & $12$ & $1.92$ & $1.48$ & $25/32$ & $29/32$ & $22/32$\\
$200$ & $14$ & $2.02$ & $1.42$ & $29/32$ & $30/32$ & $27/32$\\
$230$ & $16$ & $2.11$ & $1.41$ & $28/32$ & $32/32$ & $28/32$\\
\hline\hline
\end{tabular}
\end{center}
where $\eta$ is quoted for the post-transition slice. The progression is
informative. Lowering $\eta$ from $1.48$ to $1.42$ removes the throat filling.
Raising $c$ from $14$ to $16$ removes the remaining failures of the
post-transition slice entirely. That slice is then recovered as $(1,1,1)$ in
every sprinkling. At the best setting
\begin{equation}
\bm\beta_{\rm before}=(1,0,0)\ \longrightarrow\ \bm\beta_{\rm after}=(1,1,1)
\label{eq:fullbeta}
\end{equation}
is obtained in $28/32$ sprinklings, at about $3000\,$s per sprinkling on one
core.

All four residual failures occur on the pre-transition slice, and all are
spurious voids. They did not respond to the increase in $c$, which identifies
them as the third mode rather than the second. The cause is visible in
\eqref{eq:twoknobs}. The two slices are given the same $|L|$ and the same $c$,
but their volumes differ by a factor of about three and a half. The
pre-transition slice is therefore covered at $\eta\simeq0.93$, against
$\simeq1.41$ for its partner, and is witnessed roughly five times less densely.
It is undersampled rather than under-resolved. The remedy is a larger $N$ or a
landmark count chosen per slice, and we have pursued neither.

We therefore report \eqref{eq:dbeta} as follows. The increment $\Delta b_1=+1$
is obtained in $120$ of $120$ sprinklings across every parameter setting
examined. For a closed orientable three-manifold, Poincar\'e duality gives
$b_2=b_1$ and $b_3=b_0$, and with duality this establishes the vector. In
addition, we measure the full vector directly, without appeal to duality, in
$28/32$ sprinklings at the best setting, with the post-transition slice correct
in all of them. Duality is thus a premise for one component in a minority of
sprinklings, and a satisfied consistency condition in the rest. The detection
rates quoted here are taken from samples of at least $32$ sprinklings. Smaller
samples proved misleading: twice, a run of eight consecutive successes occurred
at settings whose rates were subsequently measured at $69\%$ and $84\%$.

The witness windows of the two slices lie entirely on their own side of the
transition. The observable therefore never evaluates a relation across it. What
is measured is the pair of slice topologies, which is exactly $\Delta\bm\beta$.
It is not, however, a test of the causal structure through the surgery. A final
geometric restriction applies: the handle must not be short and fat. For
$\lambda_{\rm th}=0.6$ with no neck, the $1$-cycle has a representative shorter than the
cover radius and is filled in, giving $0/4$. By contrast, $\lambda_{\rm th}\le0.45$ with
$\ell_{\rm neck}\gtrsim1$ is robust. The physically relevant regime, a throat
small compared with the region containing it, is the one that works.

\section{The Gauss--Bonnet term of a causal set}
\label{sec:derive}

The two curvature terms of an Einstein--Gauss--Bonnet action emerge from a causal
set by different routes. The Einstein--Hilbert term emerges from local counting.
The Benincasa--Dowker d'Alembertian~\cite{Benincasa:2010ac,Dowker:2013vba} is
built from the layers $L_k$ of elements $y$ whose order interval with $x$
contains $k-1$ elements strictly between them. In four dimensions
\begin{equation}
\mathcal{B}\phi(x)=\frac{4}{\sqrt{6}\,\ell^{2}}
\Big[-\phi(x)+\sum_{k=1}^{4}c_k\!\!\sum_{y\in L_k}\!\phi(y)\Big],
\label{eq:BD}
\end{equation}
with $c_k=(1,-9,16,-8)$, and its expectation on a sprinkling reproduces
$\int\!\sqrt{g}R$. The scalar curvature is second order in derivatives and
appears at order $\ell^{2}$ in the small-interval
expansion~\cite{Belenchia:2015aia}.

For the Gauss--Bonnet term no local construction of this kind exists, and by
Sec.~\ref{sec:kappa} none is required. The invariant $\GB$ is fourth order in
derivatives. In four dimensions $\sqrt{g}\,\GB$ is a total derivative, and the
term therefore carries no local content. A local operator for it would reproduce
a quantity that vanishes under variation. The correct discrete object is
$\kappa$. Taking the action
\begin{equation}
S=\frac{1}{16\pi G}\!\int_\Omega\!\dd^4x\sqrt{-g}\,R
+\frac{\alpha}{16\pi G}\!\int_\Omega\!\dd^4x\sqrt{-g}\,\GB+S_m ,
\label{eq:action}
\end{equation}
with $T_{\mu\nu}=-2(-g)^{-1/2}\delta S_m/\delta g^{\mu\nu}$, we find from
\eqref{eq:cgb} and \eqref{eq:localGB} that the Gauss--Bonnet term of a causal set
is
\begin{equation}
I_{GB}[C]=\frac{2\pi\alpha}{G}\sum_{v\in C}\kappa(v)
=\frac{2\pi\alpha}{G}\,\chi_\CST .
\label{eq:IGB}
\end{equation}
Adding one defect changes a count by an integer, and
\begin{equation}
\Delta I_{GB}\big|_{\text{one defect}}
=I_{GB}[C\cup\delta C]-I_{GB}[C]=\frac{2\pi\alpha}{G}\,\delta\chi_i ,
\label{eq:findiff}
\end{equation}
a difference of two integers requiring no limiting procedure. In the continuum,
background independence is imposed through a tadpole condition. Here it is the
defining property of discrete general covariance~\cite{Rideout:1999ub}, and the
Einstein--Hilbert sector contributes no dependence on $\chi$.

\subsection{The effective cosmological term from linearity in the element count}
\label{sec:Nlinear}

We now derive the effective cosmological term. The derivation is elementary, and
it forms no functional derivative at any stage.

Two ingredients are needed. The first is the causal-set relation between
cardinality and four-volume. A manifold approximates a causal set when the
latter arises from a Poisson sprinkling of unit density $\rho_c=\ell^{-4}$, which
gives
\begin{equation}
N=\rho_c V=V/\ell^{4}.
\label{eq:NV}
\end{equation}
Because the element count and the four-volume are the same quantity, any
contribution to the causal-set action proportional to the cardinality is a
cosmological constant term. The second ingredient is Eq.~\eqref{eq:split}, by
which the Euler characteristic of a region splits into a background piece and an
exactly additive sum over defects. We average the defect sum over the foam
ensemble at fixed macroscopic metric and use extensivity, $\langle N_i\rangle=n_iV$,
to obtain
\begin{equation}
\big\langle\chi\big\rangle=\chi_{\rm bg}+\Big(\dchidV\Big)V,
\qquad \dchidV=\sum_i\delta\chi_i n_i ,
\label{eq:slaved}
\end{equation}
with no boundary or corner terms, because a count of point events is additive
over any decomposition. Inserting \eqref{eq:slaved} into \eqref{eq:IGB} and using
\eqref{eq:NV},
\begin{equation}
\big\langle I_{GB}\big\rangle=\frac{2\pi\alpha}{G}\chi_{\rm bg}
\;+\;\frac{2\pi\alpha}{G}\Big(\dchidV\Big)\ell^{4}N .
\label{eq:IGBlinear}
\end{equation}
The first term is a constant. The second is linear in the cardinality of the
causal set.

That is the whole argument, because a term linear in the four-volume is what a
cosmological constant is. We write the cosmological term of \eqref{eq:action} in
the standard form $S\supset-(16\pi G)^{-1}\!\int\!\sqrt{-g}\,2\Lambda$ and use
\eqref{eq:NV}:
\begin{equation}
S_\Lambda=-\frac{\Lambda}{8\pi G}\,V=-\frac{\Lambda}{8\pi G}\,\ell^{4}N .
\label{eq:Lterm}
\end{equation}
Equations \eqref{eq:IGBlinear} and \eqref{eq:Lterm} are two contributions to the
same action. Both are linear in $N$, with all other structure identical, and
their coefficients may therefore be equated:
\begin{equation}
-\frac{\Leff}{8\pi G}=\frac{2\pi\alpha}{G}\Big(\dchidV\Big)
\quad\Longrightarrow\quad
\boxed{\;\Leff=-16\pi^{2}\alpha\,\dchidV\;}
\label{eq:Leffderived}
\end{equation}
which is precisely the expression obtained semiclassically in
Refs.~\cite{Tsilioukas:2023tdw,Anagnostopoulos:2025tax}. The field equations are
correspondingly
\begin{equation}
R_{\mu\nu}-\tfrac12 g_{\mu\nu}R+\Leff\,g_{\mu\nu}=8\pi G\,T_{\mu\nu}.
\label{eq:fieldeq}
\end{equation}
Dimensionally $[\alpha]=L^{2}$ and $[\partial\chi/\partial V]=L^{-4}$ give
$[\Leff]=L^{-2}$, as required. We also record here the dimensionless coupling
\begin{equation}
\ab\equiv\alpha/\ell_P^{2},
\label{eq:alphabar}
\end{equation}
which will carry the one free parameter of the prediction.

The derivation uses four assumptions. We list them together because the third
and fourth are usually left implicit.

\emph{(i) Extensivity}, $\langle N_i\rangle=n_iV$. This requires defects
uncorrelated beyond the mean separation, and it is what distinguishes a foam from
a manifold carrying a few handles.

\emph{(ii)} That $\partial\chi/\partial V$ is a constant fixed by the microscopic
dynamics. This is legitimate for two reasons. The nucleation rate is fixed by the
instanton action, whose dependence on the background enters only through
curvature corrections of relative order $R\ell^{2}\sim10^{-120}$. The throat size
is fixed by charge quantisation and by the decay constant, both of which are
metric independent.

\emph{(iii)} That the increments $\delta\chi_i$, computed in the Euclidean sector
where the instantons live, may be inserted into the Lorentzian effective action.
This is standard instanton logic. In the causal-set setting it is less awkward
than it appears, since by \eqref{eq:morse} the increments are Morse counts
defined order-theoretically without reference to signature.

\emph{(iv)} That the discreteness scale $\ell$ is held fixed, with the
four-volume carried by the element count. This is the causal-set postulate
$\rho_c=\ell^{-4}$. It carries substantive weight in the argument, however, and
we therefore state it explicitly. The defect content of a causal set is
$\langle\sum_i\delta\chi_i\rangle=(\sum_i\delta\chi_i\bar n_i)N$ with
$\bar n_i=n_i\ell^{4}$ dimensionless. It therefore depends on the causal set
through $N$ alone. Consider instead the alternative convention, in which $N$ is
held fixed and $\ell^{4}$ absorbs the change of four-volume. Then $n_iV=N_i$ is
invariant, the $N$-linear term carries no metric dependence, and $\Leff$
vanishes identically. The two readings are physically distinct, and the result
exists only under the first.

Equally important is what the derivation does not use. It uses no variation of
the metric, no Lanczos--Lovelock identity and no Chern--Gauss--Bonnet boundary
term. In particular, it forms no functional derivative $\delta\chi/\delta
h^{\mu\nu}$. In this respect the discrete construction differs from the
continuum one, and Sec.~\ref{sec:paschalis} makes the consequence explicit.

\subsection{The scale dependence of the topological density}
\label{sec:regimes}
\label{sec:hydro}

Although \eqref{eq:Leffderived} required no derivative, the quantity
$\partial\chi/\partial V$ that it contains is meaningful only above a certain
scale. We now locate that scale.

\emph{Coarse graining and the defect scale.} Foam defects are events, and by
Sec.~\ref{sec:kappa} the Euler density is the atomic measure \eqref{eq:atomic}.
Coarse-graining over a cell $\Omega_L$ of side $L$ and proper four-volume $V_L$
defines
\begin{equation}
\Big\langle\dchidV\Big\rangle_L\equiv\frac{1}{V_L}\int_{\Omega_L}\!\mu_\chi
=\frac{1}{V_L}\sum_{i\in\Omega_L}\delta\chi_i .
\label{eq:cgdensity}
\end{equation}
The nucleation rate is a diffeomorphism-invariant scalar density $n_i$. The count
in a cell is therefore Poisson, $\langle N_i\rangle=\mathrm{Var}(N_i)=n_iV_L$, and
\eqref{eq:cgdensity} carries a relative fluctuation
\begin{equation}
\boxed{\;\frac{\Delta\Leff}{\Leff}=\frac{1}{\sqrt{nV_L}}
=\Big(\frac{\ldef}{L}\Big)^{2},\qquad
\ldef\equiv n^{-1/4}.\;}
\label{eq:fluct}
\end{equation}
The exponent is two because the count is a four-volume.

Matching $\Lambda_{\rm obs}\ell_P^{2}=2.85\times10^{-122}$ with $|\delta\chi|=2$
and $\ab=1$ gives $n\ell_P^{4}=9.02\times10^{-125}$. Hence
$\ldef=1.03\times10^{31}\ell_P=0.166\,$mm, and there are four regimes:
\begin{center}
\begin{tabular}{lll}
\hline\hline
regime & range & density\\
\hline
sub-discreteness & $L<\ell_P$ & undefined\\
sub-window & $\ell_P<L<\lcg$ & undefined\\
atomic & $\lcg<L<\ldef$ & integer jumps\\
hydrodynamic & $L>\ldef$ & smooth\\
\hline\hline
\end{tabular}
\end{center}
with, from \eqref{eq:fluct},
\begin{center}
\begin{tabular}{lcccc}
\hline\hline
$L$ & $0.1\,$mm & $1\,$mm & $1\,$cm & $1\,$m\\
$N=nV_L$ & $0.13$ & $1.3\times10^{3}$ & $1.3\times10^{7}$ & $1.3\times10^{15}$\\
$\Delta\Leff/\Leff$ & $2.8$ & $2.8\times10^{-2}$ & $2.8\times10^{-4}$ & $2.8\times10^{-8}$\\
\hline\hline
\end{tabular}
\end{center}

The obstruction of Ref.~\cite{Paschalis:2026top} operates in the atomic regime,
and we regard that analysis as correct there. A single topological variation
produces a finite jump which does not vanish as the feature shrinks. A continuum
action defined by such variations is therefore discontinuous. Our position is not
that this conclusion is wrong. It is that, by Sec.~\ref{sec:Nlinear}, no
derivative of $\chi$ is required at any scale. The present discussion adds one
further point. If $\Leff$ is to be treated as a constant in \eqref{eq:fieldeq}
rather than as a fluctuating source, the density must also be smooth. The law of
large numbers recovers that smoothness above $\ldef$.

The scale $\ldef$ lies close to the dark-energy length scale, and this is an
identity rather than a coincidence. With $|\delta\chi|=2$ and $\ab=1$ one has
$\Lambda\ell_P^{2}=32\pi^{2}n\ell_P^{4}$, while $\rho_\Lambda=\Lambda/8\pi G$.
Therefore
\begin{equation}
\frac{\ldef}{\rho_\Lambda^{-1/4}}=\Big(\frac{32\pi^{2}}{8\pi}\Big)^{1/4}
=(4\pi)^{1/4}=1.883
\label{eq:coincidence}
\end{equation}
exactly, independently of the value of $\Lambda$. Numerically
$\ldef=0.166\,$mm against $\rho_\Lambda^{-1/4}=0.088\,$mm, the latter being the
familiar $2.24\,$meV. Each defect contributes an action increment of order unity
in Planck units. It follows necessarily that $\rho_{\rm foam}\sim n$ and
$\rho^{-1/4}\sim n^{-1/4}$. The result also has a physical reading. The
sub-millimetre scale, below which Newtonian gravity is untested, is where the
smooth description of $\Lambda$ ceases to be meaningful and the foam becomes
granular.

% ===================================================================
% FIG 3 -- the scale hierarchy: discreteness -> semiclassical -> manifold
% ===================================================================
\begin{figure}[t]
\centering
\includegraphics{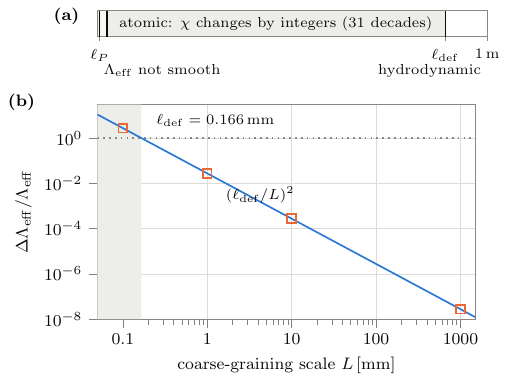}
\caption{From discreteness to a manifold. \emph{(a)} The scale hierarchy over
thirty-five decades. Below $\lcg$, in the narrow band adjoining $\ell_P$, the
spatial topology of an antichain is not defined, because the stability window of
the nerve construction has not opened. Between $\lcg$ and $\ldef=n^{-1/4}$, a
range about thirty decades wide, the Euler characteristic of a region is a sum of
integer jumps. It is not a differentiable function of the four-volume, and this
is where the continuum obstruction of Ref.~\cite{Paschalis:2026top} operates.
Above $\ldef$ it is smooth. The finite difference \eqref{eq:findiff} is exact
throughout, which is why the derivation of Sec.~\ref{sec:Nlinear} needs none of
this. Within the narrow band the two microscopic scales are ordered as the
mechanism requires, $\lcg\simeq4.35\,\ell_P<a_0\simeq6.2\,\ell_P$, and the
dominant wormhole is therefore resolvable. \emph{(b)} The relative fluctuation
\eqref{eq:fluct} of the coarse-grained topological density. Squares are the
tabulated values, and the shaded region is the atomic regime. The exponent is two
because the defect count is a four-volume. The content of
\eqref{eq:coincidence} is that $\ldef$ falls in the sub-millimetre range, where
Newtonian gravity is untested.}
\label{fig:scales}
\end{figure}

\emph{The continuum variation as a consistency check.} The derivation of
Sec.~\ref{sec:Nlinear} is self-contained. It is nevertheless instructive to see
that the continuum variational route, performed carefully, returns the same
answer, and to see exactly where the Lanczos--Lovelock identity enters. In four
dimensions the Lanczos--Lovelock tensor vanishes identically,
\begin{align}
H_{\mu\nu}=\,&2\big[RR_{\mu\nu}-2R_{\mu\alpha}R^{\alpha}{}_{\nu}
-2R_{\mu\alpha\nu\beta}R^{\alpha\beta}\nonumber\\
&+R_{\mu}{}^{\alpha\beta\gamma}R_{\nu\alpha\beta\gamma}\big]
-\tfrac12 g_{\mu\nu}\GB\;\equiv\;0
\qquad(D=4),
\label{eq:LL}
\end{align}
and hence $\delta\!\int\!\sqrt{-g}\,\GB=0$ for compactly supported $\delta g$.
Equation~\eqref{eq:LL} is an algebraic identity of the Riemann tensor. It holds
pointwise, for every metric, on every manifold, and it carries no hypothesis. It
states that for each fixed configuration the Gauss--Bonnet term contributes
nothing to the field equations. It does not state, and cannot state, anything
about topology.

The object appearing in the effective action of a foam is not
$\int\!\sqrt{-\bar g}\,\GB[\bar g]$ for the coarse-grained metric $\bar g$. It is
the ensemble average $\langle\int\!\sqrt{-g}\,\GB\rangle$ over foam
configurations. Averaging does not commute with \eqref{eq:LL}. The identity
applies configuration by configuration, whereas by \eqref{eq:slaved} the average
carries a term proportional to $V$ that no single configuration carries. That
term, and only that term, is varied. With
$\delta\!\int\!\sqrt{-\bar g}/\delta\bar g^{\mu\nu}
=-\tfrac12\sqrt{-\bar g}\,\bar g_{\mu\nu}$ one obtains \eqref{eq:Leffderived}
again. The coefficient $16\pi^{2}$ arises as $32\pi^{2}$ from \eqref{eq:cgb}
times the factor $\tfrac12$ from $\delta\sqrt{-g}$.

Two caveats attach to this route, and they are the reason we do not take it as
primary. Neither affects Sec.~\ref{sec:Nlinear}. First, the variation must be
understood as taken with respect to a metric smeared on a scale $L\gg\ldef$. A
sharp functional derivative $\delta/\delta\bar g^{\mu\nu}(x)$ is a statement at
arbitrarily fine resolution, which is precisely the atomic regime in which the
density is not smooth. The resulting equation \eqref{eq:fieldeq} is a
hydrodynamic relation between smeared quantities, as the Einstein equations of
any emergent description are. Second, one must resist the temptation to
decompose $\Omega$ into cells and apply the Chern--Gauss--Bonnet theorem cell by
cell. The Euler characteristic is not additive over such a decomposition:
$\chi$ of a four-ball cut in two is $1$ rather than $2$. The reason is that the
boundary term for a region with corners carries corner contributions that do not
cancel between neighbours. The defect count of \eqref{eq:split} is additive,
whereas $\chi$ of the cells is not. Section~\ref{sec:Nlinear} uses only the
former.

\subsection{Relation to the continuum construction}
\label{sec:paschalis}

The continuum derivation~\cite{Tsilioukas:2023tdw} uses \eqref{eq:cgb} only
through the global identity. Its one-loop apparatus exists solely to extract a
local field-equation term from a global topological quantity. In a discrete
theory that extraction is unnecessary. This bears on a rigorous obstruction.
Paschalis~\cite{Paschalis:2026top} shows that for an infinitesimal topological
variation with metric scaled as $\epsilon g'$,
\begin{equation}
\Delta S_{EGB}(\epsilon)=\epsilon\!\int_{M'}\!R[g']\sqrt{g'}\,\dd^{4}x'
+\text{const}\times\alpha\,\chi(M'),
\label{eq:paschalis}
\end{equation}
which does not vanish as $\epsilon\to0^{+}$. The action is therefore
discontinuous, and no topological functional derivative exists. The same work
further shows that the Einstein--Hilbert action admits no critical points under
topological variations in $n=4$.

The magnitude identified there is the same one appearing in the semiclassical
treatment, and indeed the same as the exact increment \eqref{eq:findiff}. What is
at issue is its interpretation as a derivative. Section~\ref{sec:Nlinear} does
not respond by negotiating the scale at which the derivative is taken. It
responds by observing that the derivation forms no functional derivative of
$\chi$ anywhere. We obtained the effective cosmological term by comparing the
coefficients of two terms linear in the cardinality of a discrete set. There is
accordingly nothing for the obstruction to obstruct. The limit
$\epsilon\to0^{+}$, in which a feature of vanishing size contributes a finite
amount, does not arise, because a discrete theory possesses a minimum four-volume
and no continuous parameter to be sent to zero. We do not claim that this
resolves every issue raised in Ref.~\cite{Paschalis:2026top}. The absence of
critical points in $n=4$ concerns the continuum variational principle, to which
we return in Sec.~\ref{sec:renorm}.

\section{Resolution, the available species, and the handle action}
\label{sec:flow}

The natural coarse-graining of a causal set is decimation. Each element is
retained with probability $f_{\rm ret}$, and the order is inherited by
restriction. A Poisson sprinkling of density $\rho_c$ decimates to one of density
$f_{\rm ret}\rho_c$, which gives
\begin{equation}
\lcg=\ell f_{\rm ret}^{-1/4}.
\label{eq:decimate}
\end{equation}
Appendix~\ref{app:thin} proves that this operation is identical to the smearing
that causal set theory already requires in order to tame the fluctuations of
\eqref{eq:BD}. Hence $\lcg$ is the nonlocality scale $\ell_k$, and no new
parameter is introduced.

\subsection{Topological features are not destroyed by decimation}
\label{sec:threshold}

\begin{figure}[t]
\centering
\includegraphics{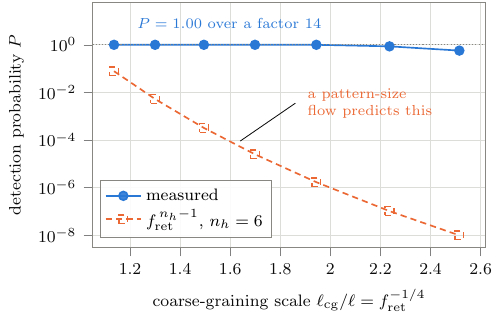}
\caption{Coarse-graining does not destroy a topological feature by attrition of
the elements that support it. A torus is sprinkled and decimated at retention
probability $f_{\rm ret}$, and the handle remains detected with probability $P$.
If survival required every one of $n_h$ supporting elements, $P$ would follow the
dashed power law. Instead it is unity across a factor of fourteen and then
collapses. A topological feature is redundantly witnessed, and it is lost only
when the resolution exceeds its physical size. At $f_{\rm ret}=0.07$ the power
law predicts $1.7\times10^{-6}$ where $1.00$ is measured. The size of the foam
features therefore controls the scale dependence of the topological density, and
no combinatorial pattern size enters: $n_h$ drops out of the physics.}
\label{fig:decimate}
\end{figure}

It might be supposed that a defect occupying $n_h$ elements survives decimation
only if all of them do. This would give $p_h\to f_{\rm ret}^{\,n_h-1}p_h$ and
hence a power-law flow. This is not the case. We sprinkle a torus, for which
$b_1=2$, and decimate it. The handle remains detected with the probability $P$
shown in Fig.~\ref{fig:decimate}. Detection is perfect across a factor of
fourteen in $f_{\rm ret}$ and then collapses. A power law with $n_h=6$ predicts
$1.7\times10^{-6}$ where $1.00$ is measured, and fitting the transition gives an
effective exponent of $0.7$. Varying the tube radius moves the failure point,
which confirms that it is set by the physical size of the handle.

The reason is that a topological feature is redundantly witnessed. A region
containing $M=(\ell_w/\ell)^{4}$ elements admits a great many sub-patterns, each
certifying the same topology. The survival probability stays near unity until the
resolution exceeds the size of the feature, and then falls sharply. The scale
dependence is therefore governed by the size of the foam features and not by any
combinatorial pattern size.

% ===================================================================
% FIG 4 -- decimation does not destroy topology by a power law
% ===================================================================

This result establishes that the naive combinatorial flow is false. Resolution
can therefore enter only through a physical size threshold. The result does not
by itself fix the density, because the spectrum of available sizes must still be
supplied. In Sec.~\ref{sec:species} we find that spectrum to be discrete, with
charge quantisation selecting a single configuration. The resolution then drops
out of the prediction altogether, and survives only as the consistency condition
$\lcg\lesssim a_0$ of Sec.~\ref{sec:pred}. That the answer is independent of the
coarse-graining scale is a result rather than an omission.

\subsection{The spectrum, the nucleation rate, and the normalisation}
\label{sec:coleman}
\label{sec:norm}

At coarse-graining scale $\lcg$ only handles of physical size
$\ell_w\gtrsim\lcg$ are resolved. Each is weighted semiclassically by
$e^{-S(\ell_w)}$, with $V/\ell_w^{4}$ available positions, and therefore
\begin{equation}
n(\ell_w)\,\dd\ln\ell_w\ \sim\ A(\ell_w)\,e^{-S(\ell_w)}\dd\ln\ell_w .
\label{eq:spectrum}
\end{equation}
Writing the resulting density as a rate per four-volume,
\begin{equation}
n_i=\Gamma_i=A_i\,e^{-\Delta I_i},
\label{eq:coleman}
\end{equation}
we recover exactly the identification of Ref.~\cite{Anagnostopoulos:2025tax}.
That identification is therefore a consequence of the present construction
rather than an additional assumption. It is the standard dilute-instanton-gas
logic of Ref.~\cite{Coleman:1988tj}.

Two candidate prefactors are available, and we carry both. The counting of
independent cells supplies $A=\ell_w^{-4}$ directly. The semiclassical
determinant supplies $A=(S/2\pi)^{n_0/2}\ell_w^{-4}$, with $n_0=4$ the number of
translational zero modes of a localised Euclidean configuration in four
dimensions. Hence $A=(S/2\pi)^{2}\ell_w^{-4}$. The two differ by
$S^{2}\sim10^{5}$. A nontrivial feature of the mechanism is that the prediction
is insensitive to the choice, at the level of three per cent in $f$, as
Sec.~\ref{sec:pred} quantifies. We do not resolve which prefactor is correct,
and the insensitivity is the reason we need not.

To evaluate handle actions we require the relation between the causal-set action
and the Einstein--Hilbert action. We write
$S_{\rm core}=N-N_1+9N_2-16N_3+8N_4$ for the combinatorial core of
\eqref{eq:BD}, and use $\langle\mathcal{B}\cdot1\rangle\to-R/2$ with
$\sum_x\to\ell^{-4}\!\int\!\sqrt{g}$. This gives
\begin{equation}
\boxed{\;S_{\rm core}=\frac{2\sqrt6\,\pi G}{\ell^{2}}\,S_{\rm EH}
=15.391\,S_{\rm EH}\quad(\ell^{2}=G).\;}
\label{eq:norm}
\end{equation}
This agrees with the causal-set action of Ref.~\cite{Pfeiffer:2023hcs}. Its
Eq.~(47), with the four-dimensional parameters $\alpha_4=-4/\sqrt6$,
$\beta_4=4/\sqrt6$ and $C_i=(1,-9,16,-8)$, gives
$\sum_x B(-2)=(8/\sqrt6\,\ell^{2})S_{\rm core}$ and hence the same factor.

A direct numerical evaluation of the handle action, by inserting a discrete
handle into a sprinkling, is not feasible at accessible sizes. There are two
reasons. First, the flat-space baseline is not zero. It ranges from $-180$ to
$-431$ for $N=300$ to $900$, growing with $N$, and these values are boundary
terms of the finite sprinkling region. Second, a handle must satisfy $a\gg\ell$
in order to be a geometry and $\text{box}\gg a$ in order to be a perturbation.
The choice $a=3\ell$ with $\text{box}=5a$ then requires
$N=15^{4}\approx5\times10^{4}$ elements, with $O(N^{2})$ interval counting.
Reference~\cite{Pfeiffer:2023hcs} reports the same barrier: causal sets large
enough for $\langle B\rangle$ to converge are too small for $\langle B^{2}\rangle$.

\subsection{The available species}
\label{sec:species}
\label{sec:round}
\label{sec:gs}

\emph{The round handle has negative action.} Consider a round
$S^{1}\!\times\!S^{3}$ handle, $\dd s^{2}=\dd\tau^{2}+a^{2}\dd\Omega_3^{2}$ with
$S^{1}$ circumference $2\pi a$. One has $R=6/a^{2}$, $\mathrm{Vol}=4\pi^{3}a^{4}$
and $\int\!\sqrt{g}R=24\pi^{3}a^{2}$, and hence
\begin{equation}
|S_{\rm EH}|=\frac{3\pi^{2}}{2}\frac{a^{2}}{G}.
\label{eq:round}
\end{equation}
The action is, however, negative. This is the conformal-factor problem of
Euclidean gravity~\cite{Gibbons:1978ac}: $e^{-S}$ would then be an enhancement,
and large handles would dominate. Pure gravity therefore does not supply a viable
foam, and matter support is required. This calculation, like that of the Nariai
sector below, concerns a closed manifold. It is therefore a constraint on
candidate species rather than an exhibition of a competing saddle. The
distinction matters, and we return to it.

\emph{The Giddings--Strominger wormhole.} Consider instead the axionic wormhole
of Giddings and Strominger~\cite{Giddings:1987cg}, with an axion of decay
constant $f$ and periodicity $2\pi$. Charge conservation gives
$a^{3}\theta'=n/(2\pi^{2}f^{2})$ with $n$ integer. Hence
$\rho=n^{2}/(8\pi^{4}f^{2}a^{6})$, and the Euclidean Friedmann equation is
\begin{equation}
(a')^{2}=1-\frac{a_0^{4}}{a^{4}},\qquad
a_0^{4}=\frac{8\pi G\,n^{2}}{24\pi^{4}f^{2}} .
\label{eq:gsfriedmann}
\end{equation}
The on-shell action follows from a single integral. With $\phi'=q/a^{3}$ and
$q^{2}=6a_0^{4}/8\pi G$,
\begin{equation}
\int\!\frac{\dd\tau}{a^{3}}=\frac{2}{a_0^{2}}\!\int_0^1\!\frac{u\,\dd u}{\sqrt{1-u^{4}}}
=\frac{\pi}{2a_0^{2}},
\end{equation}
and therefore
\begin{equation}
\boxed{\;
S=\frac{3\pi^{2}}{4}\frac{a_0^{2}}{G}
=\frac{\sqrt6\,\pi}{2}\,n\,\frac{\Mp}{f}\;}
\label{eq:gsaction}
\end{equation}
with $\Mp$ the reduced Planck mass. The coefficient is exactly half that of the
round handle \eqref{eq:round}, and the second form is the standard
Giddings--Strominger result.

Three conclusions follow. First, the sign is positive. In the scalar
formulation the on-shell action vanishes identically, since
$R=8\pi G(\partial\phi)^{2}$ makes the two terms cancel. The wormhole must
therefore be treated in the fixed-charge three-form ensemble, whose Legendre
transform yields a nonvanishing positive action. This is the form in which
wormhole effects appear throughout the axion-quality literature. Second, the
coefficient is computed rather than estimated. Third, the spectrum is
charge-quantised: $a_0\propto\sqrt n$ and $S\propto n$. The $n=1$ configuration
therefore dominates absolutely, and higher charges are suppressed by
$e^{-(n-1)S_1}$. A discrete spectrum would collapse the mechanism if $S_1$ were
of order unity, since then $\Gamma\sim\lcg^{-4}$. This does not occur, because
$S_1$ scales as $\Mp/f$ and is large for a sub-Planckian decay constant.

Unlike the closed manifolds considered above and below, the
Giddings--Strominger wormhole is an exact solution of the Euclidean field
equations possessing asymptotic ends. It is therefore a genuine handle that can
be glued into a background, which is what a foam defect must be.

\subsection{Exclusion of the $\delta\chi=+2$ sector}
\label{sec:nariai}

Equation~\eqref{eq:dchi} lists two species. By \eqref{eq:Leffderived} the sign of
$\Leff$ is the sign of $-\sum_i\delta\chi_i n_i$, and a foam dominated by
$\delta\chi=+2$ would therefore give a negative cosmological constant. The
magnitude computed above does not by itself determine the answer, and we must
examine the $\delta\chi=+2$ sector. We show that it is not available.

\emph{It is not a solution at $\Lambda=0$.} The $\delta\chi=+2$ species is
$S^{2}\!\times\!S^{2}$. For the product metric
$\dd s^{2}=a^{2}\dd\Omega_1^{2}+b^{2}\dd\Omega_2^{2}$ one has
$R_{\mu\nu}=a^{-2}g_{\mu\nu}$ on the first factor, $b^{-2}g_{\mu\nu}$ on the
second, and $R=2a^{-2}+2b^{-2}$. The only harmonic two-forms are the volume forms
of the two spheres. The most general Maxwell field therefore carries magnetic
charges $q$ and $p$ on them, and the two independent field equations are
\begin{align}
b^{-2}-\Lambda&=4\pi G\big(p^{2}b^{-4}-q^{2}a^{-4}\big),\nonumber\\
a^{-2}-\Lambda&=4\pi G\big(q^{2}a^{-4}-p^{2}b^{-4}\big),
\label{eq:s2s2blocks}
\end{align}
whose sum is flux independent:
\begin{equation}
\boxed{\;\Lambda=\tfrac12\big(a^{-2}+b^{-2}\big)>0\;}
\label{eq:s2s2trace}
\end{equation}
for every real $a$ and $b$. This is the trace equation. Maxwell stress is
traceless, which gives $R=4\Lambda$, and $R>0$ on any product of round spheres.
The axion does worse still. A nonconstant harmonic $\dd\phi$ requires $b_1\ge1$,
whereas $b_1(S^{2}\!\times\!S^{2})=0$, and the axion therefore admits no
configuration on this manifold at all. The $\delta\chi=+2$ geometry exists only
as a saddle of a theory that already possesses a positive cosmological constant.
It cannot consistently be invoked as the origin of one.

\emph{Its action is negative.} Granting \eqref{eq:s2s2trace}, the equations force
$a=b=\Lambda^{-1/2}$, the Nariai solution~\cite{Nariai:1950,Ginsparg:1982rs},
with
\begin{equation}
S_{\rm Nariai}=-\frac{2\pi}{G\Lambda}
\label{eq:nariaiaction}
\end{equation}
(the same calculation gives $-3\pi/G\Lambda$ for $S^{4}$, the standard de Sitter
value). This is the conformal-factor problem of \eqref{eq:round} in its extreme
form. For the observed $\Lambda$ the action is $-2.2\times10^{122}$, and the
$1/\Lambda$ scaling exhibits the circularity in the number itself.

\emph{Only $\delta\chi=-2$ can carry axion charge.} The two observations above
are special cases of a topological statement. For a closed orientable connected
$X$, Poincar\'e duality gives $b_3=b_1$ and $b_4=b_0=1$, and therefore
\begin{equation}
\chi(X)=2-2b_1+b_2,\qquad \delta\chi=\chi(X)-2=b_2-2b_1 ,
\label{eq:dchibetti}
\end{equation}
which reproduces \eqref{eq:dchi} and generalises it. Axion charge is
$n=\int_{S^{3}}H_3$ and therefore requires $b_3=b_1\ge1$:
\begin{center}
\begin{tabular}{lccrrc}
\hline\hline
$X$ & $b_1$ & $b_2$ & $\chi$ & $\delta\chi$ & charge\\
\hline
$S^{1}\!\times\!S^{3}$ & $1$ & $0$ & $0$ & $-2$ & yes\\
$S^{2}\!\times\!S^{2}$ & $0$ & $2$ & $4$ & $+2$ & no\\
$S^{4}$                & $0$ & $0$ & $2$ & $0$  & no\\
$\mathbb{CP}^{2}$      & $0$ & $1$ & $3$ & $+1$ & no\\
$K3$                   & $0$ & $22$& $24$& $+22$& no\\
\hline\hline
\end{tabular}
\end{center}
Three consequences follow from the table. Every simply connected defect has
$b_1=b_3=0$ and is therefore uncharged. The minimal charged defect is
$S^{1}\!\times\!S^{3}$, which is the wormhole of Sec.~\ref{sec:species}. A
charged defect with $\delta\chi\ge0$ requires $b_2\ge2b_1\ge2$. It must
therefore contain the $S^{2}\!\times\!S^{2}$ factors just shown to be
unsupportable at $\Lambda=0$. The underlying theorem runs as follows. With
$\Lambda=0$ and traceless matter, $R=0$. On a simply connected $X$ the axion is
forced constant and the solution is Ricci flat. A Ricci-flat metric on a
manifold admitting positive scalar curvature is flat~\cite{Besse:1987}. It
follows that $\chi=0$.

An unsupported handle has negative action, and pure gravity supplies no viable
foam. The supported sector is therefore exclusively $\delta\chi=-2$. Hence
$\sum_i\delta\chi_i n_i<0$ and
\begin{equation}
\Leff=-16\pi^{2}\alpha\,\dchidV>0\qquad(\alpha>0):
\label{eq:signfixed}
\end{equation}
the sign of the cosmological term is determined by the mechanism rather than
assumed. Quantitatively, with $n_i=A_ie^{-S_i}$ and comparable prefactors, the
conclusion fails only if some $\delta\chi>0$ saddle has Euclidean action below
that of the dominant wormhole. The conclusion therefore requires
\begin{equation}
S_{+}>S_{1}\simeq286 ,
\label{eq:signbound}
\end{equation}
with a $\delta\chi>0$ contamination of $e^{-(S_+-S_1)}$ when it holds.

One class of configuration is not excluded by the above, and we record it. A
compact Ricci-flat manifold solves the $\Lambda=0$ equations with vanishing
Einstein--Hilbert action. An example is $K3$, which has $\delta\chi=+22$. It is
spin with $\hat A=-\sigma/8=2$, and therefore admits no
positive-scalar-curvature metric by the Lichnerowicz
argument~\cite{Lichnerowicz:1963}. It carries no axion charge. More to the
point, a closed Ricci-flat manifold is not a defect in a background. Glued in as
a handle, it requires a neck on which the metric is not Ricci flat. The glued
configuration then does not solve the field equations, and its action does not
vanish. This is precisely the distinction drawn at the end of
Sec.~\ref{sec:species}. The status of zero-action gravitational instantons in a
Euclidean sum over topologies is a long-standing question that we do not settle
here. It is the only surviving route to a $\delta\chi>0$ species.

\emph{Self-consistency: the sector reopened by $\Leff>0$.} We established the
exclusion above at $\Lambda=0$, but the mechanism does not leave $\Lambda$
there. Equation~\eqref{eq:signfixed} generates $\Leff>0$. By
\eqref{eq:s2s2trace}, this is exactly the condition under which
$S^{2}\!\times\!S^{2}$ becomes a solution of the field equations. The very term
that the wormhole sector produces therefore reopens the channel that was closed
at $\Lambda=0$. Whether the fixed point survives is then a question about that
loop rather than about the $\Lambda=0$ theory. Three observations close it.

\emph{(i) The configuration is not a foam defect.} The Nariai solution has
$a=\Lambda^{-1/2}$. For the observed $\Lambda$ this is $9.6\times10^{25}\,$m, or
$0.70$ of the Hubble radius. The throat is $a_0=6.2\,\ell_P=1.0\times10^{-34}\,$m,
smaller by sixty orders of magnitude. A foam is a dilute gas of microscopic
defects at a fixed rate per unit four-volume, which is assumption \emph{(i)} of
Sec.~\ref{sec:Nlinear}. A single horizon-sized object is not a member of that
ensemble and cannot enter $\sum_i\delta\chi_i n_i$.

\emph{(ii) A nucleation rate is a difference of actions, not a bare action.} The
rate at which a background decays into a configuration is governed by the
difference of the two Euclidean actions. With $S_{\rm dS}=-3\pi/G\Lambda$ and
\eqref{eq:nariaiaction},
\begin{equation}
\Delta S=S_{\rm Nariai}-S_{\rm dS}=+\frac{\pi}{G\Lambda}=1.10\times10^{122},
\label{eq:gpexponent}
\end{equation}
positive and large. This is the statement of Ginsparg and
Perry~\cite{Ginsparg:1982rs} that de Sitter space is semiclassically stable
against Nariai nucleation~\cite{Bousso:1996au}. It is the reason the negative
bare action of \eqref{eq:nariaiaction} never appears in a rate.

\emph{(iii) The suppression is the horizon entropy.} Equation
\eqref{eq:gpexponent} is one third of the Gibbons--Hawking
entropy~\cite{Gibbons:1976ue} $S_{\rm dS}=3\pi/G\Lambda=3.31\times10^{122}$. A
change of topology on the horizon scale is therefore suppressed by a fixed
fraction of the horizon entropy, independently of the details of the instanton.

Hence
\begin{equation}
\frac{\Gamma_{\rm Nariai}}{\Gamma_{\rm wormhole}}\sim e^{-\Delta S+S_1}
=e^{-1.10\times10^{122}},
\label{eq:loopsupp}
\end{equation}
and the induced correction to \eqref{eq:master} is smaller than any quantity that
appears elsewhere in cosmology. The fixed point is stable, and $w=-1$ survives
the feedback.

The loop nevertheless leaves a condition, and it is not vacuous. Stability
requires $\Delta S\gg S_1$, that is
\begin{equation}
\boxed{\;G\Lambda\ \ll\ \frac{\pi}{S_1}=1.1\times10^{-2},\;}
\label{eq:selfprotect}
\end{equation}
which is satisfied by a hundred and twenty orders of magnitude. The mechanism is
in this sense self-protecting: the smaller the cosmological term it generates,
the more strongly suppressed is the channel that would remove it. It fails only
for $G\Lambda\gtrsim10^{-2}$, that is at $(G\Lambda)^{1/4}\simeq0.32\,\Mpl$. That
value belongs to an inflationary rather than a late-time era, and in such an era
the dilute-gas expansion of Sec.~\ref{sec:coleman} has broken down in any case.
Even an unsuppressed Nariai channel would not produce dynamical dark energy. It
would instead produce the relaxation of $\Lambda$ towards zero familiar from
Ref.~\cite{Coleman:1988tj}, which overshoots the observed value rather than
selecting it.

\emph{A route to $w\neq-1$ that does survive.} The exponential in
\eqref{eq:master} is a lever: a slow drift in the quantity that sets it produces
a slow drift in $\Leff$. Since $\Leff\propto e^{-S_1}$ with
$S_1=(\sqrt6\pi/2)\Mp/f$,
\begin{equation}
1+w=-\frac{S_1}{3}\,\frac{\dd\ln f}{\dd\ln a},
\label{eq:wdrift}
\end{equation}
and with $S_1\simeq286$ the gain is $S_1/3\simeq95$. A drift of one part in two
thousand per $e$-fold in the decay constant already gives $|1+w|\simeq0.05$, at
the edge of current bounds~\cite{Planck:2018vyg,DESI:2024mwx,DESI:2025zgx}. A
rising $f$ gives $w<-1$ and a falling $f$ gives $w>-1$.
Equation~\eqref{eq:wdrift} is a relation rather than a free function. It fixes
$w$ by $S_1$, which \eqref{eq:sol} already determines, together with the drift
rate of $f$. A slowly rolling decay constant is, however, a light modulus. It
inherits the usual fifth-force and equivalence-principle constraints on such
fields~\cite{Copeland:2006wr}. We therefore do not pursue it, and record only
that the scenario is not obliged to give $w=-1$ if $f$ is not strictly constant.
Of the channels we have examined, this is the only one through which the dark
energy becomes dynamical.

\subsection{The predicted decay constant, and consistency checks}
\label{sec:pred}
\label{sec:checks}

We combine \eqref{eq:Leffderived}, \eqref{eq:density} and \eqref{eq:coleman}
with $\delta\chi=-2$ from Sec.~\ref{sec:nariai} and the determinant prefactor
$A=(S/2\pi)^{2}a_0^{-4}$. This gives
\begin{equation}
\Leff=8\,\alpha\,\frac{S^{2}}{a_0^{4}}\,e^{-S}.
\label{eq:master0}
\end{equation}
Eliminating $f$ in favour of $S$ through \eqref{eq:gsaction} gives
$a_0^{2}/\ell_P^{2}=4S/3\pi^{2}=0.1351\,S$. The factor $S^{2}$ then cancels, and
\begin{equation}
\boxed{\;\Leff\,\ell_P^{2}=438.3\;\ab\;e^{-S_1},\qquad \ab=\alpha/\ell_P^{2}.\;}
\label{eq:master}
\end{equation}
The Gauss--Bonnet coupling therefore enters only through $\ln\ab$. Matching
$\Lambda_{\rm obs}\ell_P^{2}=2.85\times10^{-122}$,
\begin{align}
S_1&=286.0+\ln\ab,\qquad
\frac{a_0}{\ell_P}=\sqrt{0.1351\,S_1},\nonumber\\
f&=\frac{\sqrt6\,\pi}{2}\frac{\Mp}{S_1},
\label{eq:sol}
\end{align}
which is essentially independent of $\alpha$:
\begin{center}
\begin{tabular}{cccc}
\hline\hline
$\ab=\alpha/\ell_P^{2}$ & $S_1$ & $f\,$[GeV] & $a_0/\ell_P$\\
\hline
$10^{-4}$ & $276.7$ & $3.39\times10^{16}$ & $6.11$\\
$10^{-2}$ & $281.4$ & $3.33\times10^{16}$ & $6.17$\\
$1$       & $286.0$ & $3.28\times10^{16}$ & $6.22$\\
$10^{2}$  & $290.6$ & $3.22\times10^{16}$ & $6.27$\\
$10^{4}$  & $295.2$ & $3.17\times10^{16}$ & $6.32$\\
\hline\hline
\end{tabular}
\end{center}
An eight-decade variation of $\alpha$ shifts the predicted decay constant by about
seven per cent. For the string-inspired value $\ab=1$,
\begin{equation}
S_1\simeq286,\qquad f\simeq3.3\times10^{16}\,\mathrm{GeV},\qquad
a_0\simeq6.2\,\ell_P .
\label{eq:pred}
\end{equation}
The observed cosmological constant thus corresponds to an axion decay constant at
the grand-unification scale, within the range expected for string axions. The
$122$ orders of magnitude are not produced by a large power of a ratio. They are
produced by an exponent of order $\ln(10^{122})=281$, and the statement is that
the dominant wormhole has Euclidean action $\simeq286$.

\emph{Robustness against the prefactor.} We repeat the calculation with the
cell-counting prefactor $A=a_0^{-4}$, in which no factor $S^{2}$ appears and
hence no cancellation occurs. This gives
$\Leff\ell_P^{2}=32\pi^{2}\ab(0.1351S_1)^{-2}e^{-S_1}$, and hence $S_1=278.4$
with $f=3.37\times10^{16}\,$GeV. The shift is three per cent, despite a
prefactor differing by $S^{2}\simeq8\times10^{4}$. The exponent alone therefore
controls the prediction, which is the reason the ambiguity noted above need not
be resolved.

\emph{Resolvability.} The dominant wormhole must be resolved at the
coarse-graining scale, $\lcg\lesssim a_0$. Taking instead the round handle
\eqref{eq:round} with the resolution-selected reading gives
$\lcg\simeq4.35\,\ell_P$ for $\ab=1$, and $4.31$ to $4.38$ for $\ab=10^{-2}$ to
$10^{2}$. These values lie comfortably below $a_0=6.2\,\ell_P$. Two independent
routes therefore converge on the same few-Planck-length scale.

\emph{The Higgs triviality bound.} Since $\lcg=\ell_k$ by
Appendix~\ref{app:thin}, the bound of Ref.~\cite{deBrito:2023pli} applies. Read
literally, their Eq.~(37) gives $M_S^{\rm Upper}=(8\pi^{2}v^{2}/9D)^{1/2}$. This
yields $132\,$GeV at $\rho^{1/4}=10^{16}\,$GeV and $119\,$GeV at
$\rho^{1/4}=\Mpl$, and the latter lies below the measured $125.25\,$GeV. There
is, however, an apparent internal inconsistency in that treatment. Their
Eq.~(22) reads $m^{2}_{\rm upper}=3\lambda_0v^{2}$, and their Eq.~(36) reads
$\lambda_0<(8\pi^{2}/9)/D$. Together these give
$m^{2}_{\rm upper}<8\pi^{2}v^{2}/(3D)$, and the factor of three from Eq.~(22)
appears to have been dropped in their Eq.~(37). With it restored, the upper bound
is $229\,$GeV and $206\,$GeV respectively, and the bound is satisfied throughout
the physical range. We report both readings because the conclusion depends on
which is correct, and we do not regard the point as settled. Under the second
reading the selected scale is permitted. It then realises the confinement of the
nonlocality scale to the discreteness scale favoured in
Ref.~\cite{deBrito:2023pli} itself.

\emph{The Gauss--Bonnet self-weight.} One effect is omitted above, and it bounds
the range over which the $\alpha$-independence of \eqref{eq:master} may be
trusted. The Euclidean action of a defect contains the Gauss--Bonnet term itself.
By \eqref{eq:IGB} its magnitude is $|I_{GB}|=2\pi\ab|\delta\chi|=4\pi\ab$. The
two species of \eqref{eq:dchi} are therefore split by $e^{8\pi\ab}$ before any
gravitational difference. This reinforces the conclusion of
Sec.~\ref{sec:nariai}, in whichever direction the Euclidean continuation of the
topological term is taken. Including it in the matching replaces
$S_1=286.0+\ln\ab$ by $S_1=286.0+\ln\ab\pm4\pi\ab$. For $\ab\lesssim0.1$ the
correction is below unity and \eqref{eq:sol} stands. For $\ab\gtrsim10^{2}$ the
linear term dominates the logarithm, and the prediction is no longer independent
of $\alpha$. The insensitivity to $\alpha$ claimed above should therefore be
understood as holding for $\ab\lesssim0.1$. At the string-inspired value
$\ab=1$ the self-weight shifts $S_1$ by $\pm4\pi$ and $f$ by about five per
cent, and the last three rows of the table above should be read with that
caveat.

\section{Consequences for renormalisation and for cosmology}
\label{sec:renorm}
\label{sec:cosmo}

Before drawing those consequences, we state in one place how a process localised
at a few Planck lengths comes to control a quantity defined on the Hubble scale.
The two scales are separated by some sixty decades, and the connection is easily
mistaken for an extrapolation. It is not one. The mechanism involves no
propagation of a Planckian fluctuation out to the horizon, and no running of a
coupling between the two scales. What it involves is extensivity, and the
argument has three steps.

First, the microphysics fixes a \emph{rate}, not a length. The dominant defect is
the wormhole of Sec.~\ref{sec:species}. Its size $a_0$ and Euclidean action
$S_1$ are determined by charge quantisation and the axion decay constant, with no
reference to any cosmological quantity. This is why \eqref{eq:gsaction} to
\eqref{eq:master} contain no epoch-dependent parameter. The semiclassical weight
$e^{-S_1}$ converts that action into a nucleation probability per unit
four-volume, Eq.~\eqref{eq:coleman}, which is a diffeomorphism-invariant scalar
density. A rate per unit four-volume has the same numerical value whether it is
evaluated in a Planckian cell or in a Hubble volume.

Second, extensivity turns the rate into a count. Assumption \emph{(i)} of
Sec.~\ref{sec:Nlinear} states that the defects are uncorrelated beyond their
mean separation. The expected number in a region is therefore $\langle
N_i\rangle=n_iV$, strictly proportional to the four-volume. The Euler
characteristic of that region is correspondingly the background value plus an
additive integer count, Eq.~\eqref{eq:slaved}. Nothing here is a long-wavelength
mode. The defects remain microscopic at every epoch, and it is their number, not
their size, that grows with the region.

Third, by Eq.~\eqref{eq:Lterm}, a term linear in the four-volume is a
cosmological constant term. The Planckian scale therefore enters the infrared
only through the dimensionless exponent $S_1$ in \eqref{eq:master}. The sixty
decades separating $a_0$ from $\Lambda^{-1/2}$ are not produced by any dynamical
transport between the scales. They are produced by the exponential $e^{-S_1}$,
with $S_1$ of order a few hundred. This is also why the prediction is
insensitive to the coarse-graining scale $\lcg$. By Sec.~\ref{sec:threshold} the
resolution enters only through which defects are counted, and charge
quantisation leaves a single species to count.

The scale $\ldef=n^{-1/4}$ of Eq.~\eqref{eq:fluct} is the one place where an
intermediate length appears. It marks a change in the \emph{description} rather
than in the physics. Below it the topological density is a sum of integer jumps,
and above it, by the law of large numbers, a smooth function. The relation
$V\propto H^{-4}$ invoked in Sec.~\ref{sec:fluct} is a statement about the past
four-volume of a single-component era. It enters only the variance, never the
mean, which is constant by construction.

The one-loop counterterm basis $\{R^{2},R_{\mu\nu}R^{\mu\nu},
R_{\mu\nu\rho\sigma}R^{\mu\nu\rho\sigma}\}$ reduces by
\begin{equation}
R_{\mu\nu\rho\sigma}R^{\mu\nu\rho\sigma}=\GB+4R_{\mu\nu}R^{\mu\nu}-R^{2},
\label{eq:GBidentity}
\end{equation}
provided that $\int\!\sqrt{g}\,\GB$ is constant. With this reduction, the
one-loop divergence of pure gravity becomes proportional to the field equations
and vanishes on shell. Pure gravity is then one-loop
finite~\cite{tHooft:1974toh}, although the finiteness does not survive two
loops~\cite{Goroff:1985th}. If the topology fluctuates, the reduction fails. By
\eqref{eq:cgb} the Gauss--Bonnet integral is $32\pi^{2}\chi$, which is now
dynamical, and $R_{\mu\nu\rho\sigma}R^{\mu\nu\rho\sigma}$ becomes independent.
One-loop finiteness therefore holds only insofar as topology change is
negligible. By \eqref{eq:master} the violation is of relative order
$e^{-S_1}=6.5\times10^{-125}/\ab$.

The topological sector is in this sense exponentially irrelevant. Writing
$\hat\Lambda\equiv\Leff\ell_P^{2}$, Eq.~\eqref{eq:master} gives
\begin{equation}
\hat\Lambda(\ab,S_1)=438.3\,\ab\,e^{-S_1},
\end{equation}
and since $S_1\propto a_0^{2}$, the topological contribution is suppressed
exponentially in the squared size of the dominant configuration rather than by
any power of it. This one fact accounts for three things at once: the emergence
of a smooth manifold, the guarantee that topology change does not spoil
low-energy predictivity, and the reason that foam is unobservable.

The comparison with the continuum treatment is instructive here as well. The
obstruction of Ref.~\cite{Paschalis:2026top} concerns well-posedness: without
critical points the stationary-phase principle fails. That analysis finds the
pathology cured for $n>4$. Within the continuum framework, extra dimensions are
therefore required. Causal set theory offers an alternative at $n=4$. Its path
integral is a discrete sum requiring no stationary phase, and the variation
\eqref{eq:findiff} is a finite difference. By Sec.~\ref{sec:Nlinear}, the
effective cosmological term is obtained without any variational principle at all.
Whether the discrete sum possesses a good classical limit is a separate open
question.

\subsection{The equation of state}
\label{sec:eos}

Equations \eqref{eq:gsaction} to \eqref{eq:master} contain no cosmological
quantity: $S_1$ is fixed by $f$, and $a_0$ by $S_1$. This rests on
manifoldlikeness. For a sprinkling into a smooth spacetime the local interval
abundances are universal, with curvature corrections of relative order
$R\ell^{2}\sim10^{-120}$. The nucleation rate per unit four-volume is therefore
the same at every epoch. Hence
\begin{equation}
\Leff=\text{const}\quad\Longrightarrow\quad w=-1
\label{eq:w}
\end{equation}
exactly. The scenario predicts a genuine cosmological constant rather than a
dynamical dark-energy sector. In this respect it is not distinguishable from
$\Lambda$CDM by background observables, and what it offers instead is a value.
The one way in which this conclusion can fail is a decay constant that is not
strictly constant, and Eq.~\eqref{eq:wdrift} quantifies that case.

The confrontation with data is correspondingly constrained. Since
\eqref{eq:w} is exact, the background expansion history is that of $\Lambda$CDM
with no additional parameter. The scenario therefore inherits, rather than
modifies, the fits to the cosmic microwave background~\cite{Planck:2018vyg}, to
baryon acoustic oscillations~\cite{DESI:2024mwx,DESI:2025zgx} and to the
supernova Hubble diagram. A Markov chain sampling of the present model over
those datasets would recover the $\Lambda$CDM posterior identically. The content
of the prediction lies in the value of $\Leff$ rather than in its evolution, and
background observables therefore cannot test it at all. The single exception is
Eq.~\eqref{eq:wdrift}. It is a one-parameter extension in the logarithmic drift
rate of the decay constant, and it maps onto the usual $w_0$ and $w_a$
parameterisation. Given the lever $S_1/3$, present bounds on $w$ already
constrain that drift rate to be below roughly one part in $10^{3}$ per $e$-fold.
Any future determination of $w\neq-1$ would convert \eqref{eq:wdrift} from a
caveat into a measurement of the time dependence of $f$. We regard that as the
only avenue by which the mechanism could be tested against expansion-history
data.

\subsection{The fluctuation channel, and its relation to everpresent $\Lambda$}
\label{sec:fluct}

The causal set is a Poisson process. The topological charge in a region
therefore has both a mean and a variance, and the two are physically distinct
channels.

The mean is the subject of this paper, and it gives \eqref{eq:w}. The variance is
\begin{equation}
\sigma_{\Leff}=16\pi^{2}\alpha\Big[V^{-1}\sum_i\delta\chi_i^{2}\Gamma_i\Big]^{1/2},
\label{eq:sigma}
\end{equation}
In any single-component era $V\propto H^{-4}$, since both the past four-volume
and $H^{-4}$ scale as $t^{4}$ for $a\propto t^{n}$. Hence
$\sigma_{\Leff}\propto H^{2}$. This reproduces the everpresent scaling of
Refs.~\cite{Sorkin:1990bh,Ahmed:2002mj} for the $\delta\chi\neq0$ sector, as it
must. The everpresent mechanism is the statement that a fluctuating four-volume
drives a fluctuating $\Lambda$ of magnitude $|\Lambda|\sim1/\sqrt V$. The
topology-change events counted here are one realisation of that fluctuation.

The two channels are therefore related not by competition but by order of
moment, and this is the sense in which the present mechanism is new. The
everpresent programme takes the mean of the topological charge to vanish, as it
must if species of both signs nucleate at comparable rates.
Appendix~\ref{app:parity} shows that order reversal does not by itself enforce
such a cancellation. The cancellation reduces instead to the dynamical condition
$n_2=n_1+n_3$. Section~\ref{sec:nariai} shows that this condition fails, and
fails badly. The $\lambda=2$ species is not supported at all, which gives
$n_2=0$. The mean therefore does not cancel, and a constant term survives
alongside the fluctuation.

Nothing here bears on the viability of the fluctuation channel itself. The
cosmology of stochastic everpresent models, in which $\Lambda$ changes sign on
roughly a Hubble time, has been studied in detail~\cite{Zwane:2017xbg}. It is a
separate question with its own literature, and we do not settle it here. We
observe only that a naive deterministic reading of $\sigma_\Lambda\propto H^{2}$
is excluded. That reading takes $\Omega_{\rm DE}$ to be a fixed fraction at every
epoch. It therefore makes the dark sector track the dominant component, giving
$w=0$ in matter domination and no acceleration at all. It also leaves a constant
$\Omega_{\rm DE}\simeq0.7$ at nucleosynthesis, corresponding to
$\Delta N_{\rm eff}\simeq17$ against a bound of order $0.4$. The stochastic
models are not subject to this argument. Our mechanism does not depend on the
outcome either way, because the mean and the variance add.

\section{Discussion}
\label{sec:disc}

We have investigated the coupling between topology change and the Gauss--Bonnet
term within causal set theory.

The Euler density of a discrete space is an exact, local, purely combinatorial
object. The quantity $\kappa$ of Eq.~\eqref{eq:kappa} sums identically to
$\chi$, and it coincides with the Regge deficit under the conditions identified
in Sec.~\ref{sec:regge}. It vanishes pointwise on flat manifoldlike regions in
two and four dimensions, and it is supported atomically on defects. The
Gauss--Bonnet term \eqref{eq:IGB} is therefore determined rather than postulated.

That term is linear in the element count, and in causal set theory the element
count is the four-volume. It is therefore a cosmological constant term, whose
coefficient can be read off by comparison:
$\Leff=-16\pi^{2}\alpha\,\partial\chi/\partial V$. No functional derivative is
formed at any stage. In this respect the discrete construction differs from the
continuum one, and it disposes of the obstruction of
Ref.~\cite{Paschalis:2026top} rather than negotiating with it.

Causal set theory already predicts a cosmological term as a fluctuation, and that
prediction takes the mean of the topological charge to vanish. It does not
vanish. Of the two topology-change species, $S^{2}\!\times\!S^{2}$ solves the
Euclidean field equations only for $\Lambda>0$. The flux terms cancel
identically in the trace, and no Maxwell configuration therefore supports it at
$\Lambda=0$. Moreover, no simply connected defect can carry axion charge, since
$b_3=b_1=0$. The supported sector is exclusively $\delta\chi=-2$, and the sign of
$\Leff$ is therefore derived rather than assumed.

The resulting effective term, Eq.~\eqref{eq:master}, gives $S_1\simeq286$,
$a_0\simeq6.2\,\ell_P$ and $f\simeq3.3\times10^{16}\,$GeV. These values are
stable to three per cent against a prefactor ambiguity of five orders of
magnitude. They are stable to about seven per cent under an eight-decade variation of
$\alpha$ when the Gauss--Bonnet self-weight is neglected. That neglect is
justified for $\ab\lesssim0.1$, and at $\ab=1$ the self-weight shifts $f$ by
about five per cent. The dominant wormhole is resolvable at the
coarse-graining scale, which by Appendix~\ref{app:thin} is itself the
nonlocality scale. Manifoldlikeness gives $w=-1$ exactly.

\subsection{What is and is not established}
\label{sec:established}

We now separate the parts of this work that are internal to causal set theory
from those imported from continuum semiclassical gravity. The distinction is not
always obvious, and a reader is entitled to it.

Five parts are internal to causal set theory:
the Euler density $\kappa$ and its properties (Sec.~\ref{sec:kappa}); the
numerical demonstrations in $2+1$ and $3+1$ dimensions
(Secs.~\ref{sec:numerics} and \ref{sec:numerics4}); the Benincasa--Dowker
normalisation \eqref{eq:norm}; the identification of decimation with smearing
(Appendix~\ref{app:thin}); and the derivation of \eqref{eq:Leffderived} from
linearity in $N$, which uses \eqref{eq:NV} essentially.

Imported are the instanton action \eqref{eq:gsaction}, the prefactor, the
nucleation rate \eqref{eq:coleman}, and hence the numerical value of $f$. Causal
set theory has no established quantum dynamics, and the rate that sets the
magnitude comes from outside the theory. The claim of this paper is accordingly
not that $\Lambda$ is derived within causal set theory. The claim is that the
Euler term is exactly and locally defined in a discrete setting, that it yields a
cosmological constant without any functional derivative, and that its sign is
fixed. The magnitude then follows from a single semiclassical input.

Three further limitations remain. First, the quantity $\chi_\CST$ of
\eqref{eq:IGB} is defined but not directly computable on a causal set, because
the order-theoretic complex of Sec.~\ref{sec:complex} is an open problem. The
increments $\delta\chi_i$ reach us through the Morse bridge \eqref{eq:morse},
whose identification of indices from order data is necessary but not sufficient
(Sec.~\ref{sec:betti}). Second, manifoldlikeness is assumed throughout and is
the premise of $w=-1$. The overwhelming majority of causal sets are not
manifoldlike, and causal set theory cannot presently discharge this assumption.
Third, the $3+1$-dimensional demonstration of Sec.~\ref{sec:numerics4} rests on
$\Delta b_1=+1$, obtained in $120$ of $120$ sprinklings, together with
Poincar\'e duality. The full Betti vector is measured directly in $28$ of $32$
sprinklings at the best resolution examined. The residual is an undersampling of
the pre-transition slice rather than a failure of the method. Moreover, the
witness windows of the two slices lie on their own sides of the transition. The
observable therefore establishes the pair of slice topologies, not the causal
structure through the surgery itself.

We close with the principal open problems. \emph{(i)} The status of Euclidean
axion wormholes as saddles of the Lorentzian theory, on which the positivity of
\eqref{eq:gsaction} depends. \emph{(ii)} Two residuals of the numerical
demonstration. One is a landmark count chosen separately for each slice, or a
larger $N$, which should remove the undersampling of the pre-transition slice
that accounts for every remaining failure in Sec.~\ref{sec:numerics4}. The other
is an observable that evaluates the causal structure across the transition
rather than the slice topologies on either side of it. Separately, a direct
causal-set evaluation of the handle action is still needed. It requires
$N\gtrsim5\times10^{4}$ and control of the boundary terms discussed in
Sec.~\ref{sec:coleman}. \emph{(iii)} The order-theoretic complex of
Sec.~\ref{sec:complex}, and with it a direct evaluation of $\kappa$ on a causal
set. \emph{(iv)} Whether the predicted $f$ is compatible with an independent
determination of the axion sector, which would turn \eqref{eq:pred} into a test.
\emph{(v)} The role of zero-action Ricci-flat gravitational instantons in the sum
over topologies, the only surviving route to a $\delta\chi>0$ species
(Sec.~\ref{sec:nariai}). \emph{(vi)} The Euclidean continuation of the
topological term, which fixes the direction of the Gauss--Bonnet self-weight
discussed in Sec.~\ref{sec:pred}.

\begin{acknowledgments}
The author would like to acknowledge the
contribution of the LISA CosWG, and of COST Actions CA21136 ``Addressing
observational tensions in cosmology with systematics and fundamental physics
(CosmoVerse)'', CA21106 ``COSMIC WISPers in the Dark Universe: Theory,
astrophysics and experiments (CosmicWISPers)'', and CA23130 ``Bridging high and
low energies in search of quantum gravity (BridgeQG)''.
\end{acknowledgments}

\section*{Code availability}

The code supporting the numerical results of
Secs.~\ref{sec:numerics} and \ref{sec:numerics4} is openly available at
\url{https://github.com/stelios-tsilioukas/topology-change-in-causal-set-theory},
together with the checks underlying the analytic claims. It is archived at
\href{https://doi.org/10.5281/zenodo.23611060}%
{doi:10.5281/zenodo.23611060}, which identifies the release used here.

\appendix

\section{Parity of $\delta\chi$ under order reversal}
\label{app:parity}

Order reversal, $x\preceq^{*}y\Leftrightarrow y\preceq x$, might be expected to
exchange the species \eqref{eq:dchi} and to enforce cancellation of the mean. It
does not, because $\chi$ is invariant. \emph{(a)} Order reversal changes the
causal orientation, not the underlying topological space, and $\chi$ is a
homotopy invariant. \emph{(b)} The chains of a poset coincide setwise with those
of its dual, and hence $\Delta(P)=\Delta(P^{*})$. Numerically, on
transitive-percolation causal sets with $n=12$ to $24$, the $j$-chain count
vectors agree term by term. \emph{(c)} Order reversal sends $f\to-f$ and maps
index $\lambda$ to $n-\lambda$. It therefore sends
$(-1)^{\lambda}\to(-1)^{n}(-1)^{\lambda}$, which is invariant precisely when $n$
is even.

The handle decomposition makes this explicit:
$\delta\chi=-2=(-1)^{1}+(-1)^{3}$ and $\delta\chi=+2=(-1)^{2}+(-1)^{2}$. A
wormhole therefore consists of one index-1 and one index-3 critical point, and a
Nariai instanton of two index-2 points. Order reversal exchanges
$\lambda=1\leftrightarrow3$ and fixes $\lambda=2$: each species maps to itself.
Cancellation of the mean would therefore reduce to the dynamical condition
$n_2=n_1+n_3$. By Sec.~\ref{sec:nariai} the $\lambda=2$ species is not supported,
which gives $n_2=0$. No cancellation occurs, and the mean survives with a
definite sign.

\section{Decimation equals smearing: a binomial thinning theorem}
\label{app:thin}

\emph{Claim.} Averaged over decimations at retention probability $\varepsilon$,
the sharp Benincasa--Dowker operator equals the operator smeared with the
standard four-dimensional weight. Hence $\varepsilon=f_{\rm ret}$ and
$\ell_k=\lcg$.

\emph{Proof.} Decimation retains each element independently with probability
$\varepsilon$. An order interval of cardinality $n$ is therefore mapped to one of
cardinality $k$ with probability
$\binom{n}{k}\varepsilon^{k}(1-\varepsilon)^{n-k}$, which is binomial thinning.
The expected layer sum is then
\begin{align}
\sum_{k=0}^{3}c_k\binom{n}{k}\varepsilon^{k}(1-\varepsilon)^{n-k}
&=(1-\varepsilon)^{n}\Big[1-9nr\nonumber\\
&\quad+16\tbinom{n}{2}r^{2}-8\tbinom{n}{3}r^{3}\Big],
\end{align}
with $r=\varepsilon/(1-\varepsilon)$ and $c_k=(1,-9,16,-8)$. Using
$16\binom{n}{2}=8n(n-1)$ and $8\binom{n}{3}=\tfrac43n(n-1)(n-2)$, this is
identically the smearing weight
\begin{align}
f_4(n,\varepsilon)=(1-\varepsilon)^{n}\big[&1-9nr+8n(n{-}1)r^{2}\nonumber\\
&-\tfrac43 n(n{-}1)(n{-}2)r^{3}\big]
\label{eq:f4}
\end{align}
of Refs.~\cite{Sorkin:2007qi,Aslanbeigi:2014zva}. Including $\varepsilon^{2}$
for the survival of the two endpoints and $\varepsilon$ for the element count, we
obtain $\langle S^{\rm sharp}[C_\varepsilon]\rangle
=\varepsilon\,S^{\rm smeared}_{\varepsilon}[C]$.

We have verified the identity to machine precision for
$\varepsilon\in\{0.05,0.2,0.5,0.8\}$ and $n\le17$. We have also confirmed the
end-to-end relation on a sprinkling of $N=900$ elements into a four-dimensional
causal diamond.

\bibliography{cst_tde}

\end{document}